\documentclass[aps,prd,reprint,superscriptaddress,nofootinbib,longbibliography,floatfix]{revtex4-2}

\usepackage[T1]{fontenc}
\usepackage{lmodern}
\usepackage[utf8]{inputenc}
\usepackage{amsmath,amssymb,bm}
\usepackage{graphicx}
\usepackage{booktabs}
\usepackage{microtype}
\usepackage{xcolor}
\usepackage{tikz}
\usetikzlibrary{arrows.meta,positioning}
\usepackage{hyperref}
\usepackage{orcidlink}
\usepackage{placeins}
\hypersetup{
  colorlinks=true,
  linkcolor=blue,
  citecolor=blue,
  urlcolor=blue,
}
\allowdisplaybreaks
\graphicspath{{figures/}}

\newcommand{\dif}{\ensuremath{\mathrm{d}}}
\newcommand{\imag}{\ensuremath{\mathrm{i}}}
\newcommand{\ee}{\mathrm{e}}
\newcommand{\N}{\mathcal{N}}
\newcommand{\D}{\mathcal{D}}
\newcommand{\Pp}{\mathcal{P}}
\newcommand{\rs}{r_{\rm s}}
\newcommand{\wtil}{\widetilde{\omega}}
\newcommand{\kb}{\bar\kappa}
\newcommand{\PhiQ}{\Phi_{0}}
\newcommand{\Order}{\mathcal{O}}
\renewcommand{\Re}{\operatorname{Re}}
\renewcommand{\Im}{\operatorname{Im}}
\definecolor{revisionblue}{RGB}{0,0,0}

\begin{document}

\title{Josephson interferometry in an Oppenheimer--Snyder-like scale-dependent black-hole spacetime}

\author{Ali \"Ovg\"un \orcidlink{0000-0002-9889-342X}}
\email{ali.ovgun@emu.edu.tr}
\affiliation{Physics Department, Faculty of Arts and Sciences, Eastern Mediterranean University, Famagusta, 99628 North Cyprus via Mersin 10, T\"urkiye}

\author{Reggie C. Pantig \orcidlink{0000-0002-3101-8591}}
\email{rcpantig@mapua.edu.ph}
\affiliation{Physics Department, School of Foundational Studies and Education, Map\'ua University, 658 Muralla St., Intramuros, Manila 1002, Philippines}

\author{G.~Lambiase \orcidlink{0000-0001-7574-2330}}
\email{lambiase@sa.infn.it}
\affiliation{Dipartimento di Fisica ``E.R. Caianiello'', Universit\`a degli Studi di Salerno, Via Giovanni Paolo II, 132, 84084 Fisciano (SA), Italy}
\affiliation{Istituto Nazionale di Fisica Nucleare, Gruppo Collegato di Salerno, Sezione di Napoli, Via Giovanni Paolo II, 132, 84084 Fisciano (SA), Italy}

\begin{abstract}
We develop a covariant framework for Josephson transport, superconducting interference, and shunt noise in the scale-dependent exterior generated by an Oppenheimer--Snyder-like collapse. Static Josephson frequencies and transported currents referred to Killing time acquire one lapse factor, whereas power acquires two. For a junction comoving with the collapsing surface, the coordinate-time phase rate differs from the frequency received at infinity because of null propagation and Doppler effects. We derive the exact radial-null travel-time kernel and show that its local near-extremal logarithmic enhancement crosses over, at fixed emission offset, to pole-controlled extremal behavior. We also obtain the redshifted resistively and capacitively shunted-junction equation and the dc and microwave-driven two-junction interference envelopes. In the negligible-total-inductance limit, static lapse imbalance modifies lobe amplitudes without shifting their centers, whereas microwave-induced translations require dynamical fluxoid closure. In a Hartle--Hawking state, Tolman redshift produces a superconducting exclusion layer and a lapse-independent asymptotic shunt-noise spectrum; its low-frequency limit obeys a parameter-free fluctuation--delay relation. Finally, we derive a shadow--Josephson consistency relation and sensitivity bounds on the running parameter.
\end{abstract}


\maketitle

\section{Introduction}\label{sec:introduction}

The Josephson effect translates a gauge-invariant condensate phase into a current and a frequency standard. For a weak link, the local current--phase relation is $I=I_c\sin\varphi$, whereas an electrochemical-potential difference drives $\dif\varphi/\dif\tau=2e\,\Delta V^{\rm p}/\hbar$ \cite{Josephson:1962zz,tinkham1975introduction,Barone_1982}. Microwave locking produces the Shapiro spectrum, and a two-junction superconducting loop converts the phase constraint into the flux-periodic response of a direct-current superconducting quantum interference device (dc SQUID) \cite{Shapiro:1963nhj,Langenberg_1966,Jaklevic:1964ysq,Clarke_2004}. Because the Josephson conversion factor is fixed by fundamental constants, the effect provides a natural probe of how local clocks, transported charge, and electromagnetic phases are compared in curved spacetime \cite{Tiesinga_2021}.

The interaction of superconducting phases with gravity has been studied since the covariant analyses of DeWitt and Anandan \cite{DeWitt:1966yi,Anandan:1981zd,Anandan_1984}. Covariant generalizations of thermoelectromagnetic and Josephson effects in superconductors, together with early superconducting tests of relativistic gravity, were developed in~\cite{Anandan:1984rte,Jain:1987rel}. Relativistic conduction, thermoelectric response, and gravitoelectromagnetic transport have subsequently been developed for conducting and superconducting media \cite{Ahmedov:1998mif,Ahmedov:1999bqr,Ahmedov:2002bx,Ahmedov:2010mt,Gavassino:2025bnx}. Weak-field studies have examined gravity-induced alternating-current (ac) Josephson oscillations, vortex configurations, and effective gravitoelectromagnetic couplings \cite{Ummarino:2020loo,Ummarino:2021tpz,Ummarino:2021vwc,Das:2023qqx,Gallerati:2022pgh,Gallerati:2022nwm}. More recent analyses have addressed Josephson transport in Schwarzschild backgrounds, including plasma effects, and in slowly rotating spacetimes \cite{Pantig:2025,AYTIMBETOV2026140781,Nurmukhammed:2026nil,Aytimbetov:2026bth}. Related developments include holographic Josephson junctions and transport \cite{Horowitz:2011dz,Ovgun:2026yeo,Santos:2024cwf,Santos:2025ugv}, wormhole-supported Josephson systems \cite{Takeuchi:2021kjv}, and curved-metric or analogue implementations in superconducting media and SQUID arrays \cite{Rastkhadiv:2025ibd,Maceda:2025jti}. Complementary holographic-superconductor models with backreaction, higher-curvature or higher-derivative couplings, and nonlinear electrodynamics illustrate how bulk gravitational and gauge dynamics can alter condensation and transport \cite{Wang:2016jov,Kuang:2011dy,Kuang:2013oqa,Kuang:2010jc,Sheykhi:2018mzs,Gangopadhyay:2012gx,Banerjee:2012vk}.

A black-hole calculation requires more than inserting a redshift factor into a flat-space formula. One must specify the observer's clock, the hypersurface through which charge is counted, and the protocol used to transmit a voltage or microwave phase. In a stationary spacetime, the Killing field converts proper-time phase evolution into asymptotic time, whereas the $3+1$ continuity equation determines the charge crossing a surface per unit Killing time \cite{Thorne_1982,Gourgoulhon:2012ffd,Poisson:2009pwt}. Spatially separated terminals additionally accumulate a propagation phase governed by the null travel time. In the weak-field regime this is the Shapiro time delay, while near a horizon its logarithmic enhancement is encoded in the tortoise coordinate and controlled by the surface gravity \cite{Shapiro:1964uw,Possel:2019zbu}. Thermal equilibrium is independently constrained by the Tolman--Ehrenfest law \cite{Tolman:1930zza,Tolman:1930ona,Lima:2019brf}. Keeping clock conversion, current transport, propagation delay, acceleration, and temperature conceptually distinct is essential because these quantities exhibit different near-horizon scalings. In a recent Schwarzschild analysis, this operational separation led to a one-lapse transfer law for the ac Josephson frequency and the asymptotic critical current of a static junction \cite{Pantig:2025}.

The one-lapse Josephson dictionary and the local weak-link baseline build directly on Ref.~\cite{Pantig:2025}.  The additions specific to the present work are the analytic scale-dependent horizon and throat structure, the collapsing-worldline distinction between coordinate and received signals, the exact rational null kernel including its extremal pole, the lapse-transformed dynamics of a resistively and capacitively shunted junction (RCSJ), the Hartle--Hawking proper-distance bound, the Tolman--KMS mapping of equilibrium shunt noise and its fluctuation--delay relation, and the shadow--Josephson consistency relation.  The microwave two-arm result below is an extension only after the time-dependent fluxoid constraint is imposed; it is not an independent static-flux effect.

These distinctions are increasingly relevant to quantum metrology. Matter-wave interferometry has directly observed a gravitational Aharonov--Bohm phase, while complementary analyses have clarified its relation to gravitationally induced proper-time and phase differences \cite{Overstreet:2021hea,Chiao:2023ezj}. Atomic clocks and atom interferometers now provide controlled tests of gravitational redshift over laboratory baselines, including miniature clock networks \cite{Pumpo:2021sok,Zheng:2022hwj}. Cryogenic torsion-balance measurements have also placed direct bounds on possible equivalence-principle violations involving superconducting niobium and the Cooper-pair sector \cite{Ross:2025wep}. The gravitational metrological triangle proposes a direct consistency comparison among clock, atom-interferometric, and superconducting standards \cite{Lammerzahl2024}. More broadly, massive quantum systems are being developed as interfaces between quantum mechanics and gravity, and gravitational time dilation can be interpreted operationally as a dephasing channel for spatially distributed quantum systems \cite{Bose:2024nhv,Balatsky:2024tbv}. Josephson devices are especially attractive in this context because phase evolution, charge transport, and microwave locking probe distinct operational manifestations of spacetime geometry.

Scale-dependent gravity provides a phenomenological framework, motivated in part by the asymptotic-safety program, in which short-distance gravitational corrections are encoded in running couplings \cite{Reuter:2012qeg}. The renormalization-group-improved black-hole prescription underlying the running Newton coupling was developed in Ref.~\cite{Bonanno:2000}. Connections between asymptotically safe gravity and generalized-uncertainty-principle phenomenology have been investigated \cite{Lambiase:2022}, and scale-dependent couplings have also been employed to construct Planck-star configurations \cite{Scardigli:2023}. Here we apply the covariant Josephson framework to the exterior generated by the Oppenheimer--Snyder-like collapse model of Ref.~\cite{Hassannejad:2024cbu}. Depending on the running parameter, the geometry contains two horizons and a curvature-finite but generally nonanalytic center, an extremal horizon with a long throat, a horizonless branch, or a finite-radius singular shell hidden inside one outer horizon. The collapse model additionally provides a physically distinguished radial geodesic, namely the stellar surface, and therefore permits a clean comparison between a junction supported at fixed areal radius and a junction comoving with the collapsing matter.

The analysis yields seven principal results.  First, the horizon, singular-shell, extremal, and finite-curvature core structure is obtained analytically for $\gamma>0$.  Second, static Josephson frequency and current observables carry one lapse factor, whereas asymptotic power carries two.  Third, for a collapsing terminal we distinguish the coordinate-time phase rate from the outgoing signal received at infinity; both scale as $F$ near a simple horizon, but differ by a finite Doppler factor and by their natural time coordinate.  Fourth, the exact radial-null travel-time kernel, its nonuniform simple-horizon-to-extremal crossover, and its double-pole extremal limit are derived.  Fifth, the RCSJ equation and the dc and dynamically closed microwave-driven SQUID envelopes are obtained.  Sixth, a Hartle--Hawking bath produces a state-qualified superconducting exclusion layer and a lapse-independent, asymptotic-time-referred shunt-noise spectrum whose low-frequency limit obeys a parameter-free fluctuation--delay relation.  Seventh, the leading shadow and one-lapse shifts are combined into a parameter-eliminated cross-channel consistency relation.

The condensate and electromagnetic field are treated as test systems on a prescribed background. The near-horizon formulas should therefore be interpreted as invariant scaling relations and as targets for analogue or thought experiments, rather than as a claim that an ordinary laboratory device can be stationed arbitrarily close to an astrophysical horizon.

Section~\ref{sec:geometry} analyzes the scale-dependent exterior. Section~\ref{sec:covariant} develops the covariant Josephson dictionary. Section~\ref{sec:ac} treats microwave locking and null propagation. Section~\ref{sec:collapse} distinguishes coordinate and received signals from a collapsing terminal. Section~\ref{sec:dc} derives the local weak-link and RCSJ equations. Section~\ref{sec:squid} studies dc and dynamically closed microwave-driven SQUID interference. Section~\ref{sec:noise-delay} develops the Tolman--KMS shunt-noise spectrum and the horizon fluctuation--delay relation. Section~\ref{sec:validity} discusses support, curvature, thermal constraints, and observability. Section~\ref{sec:shadow-consistency} derives the shadow--Josephson consistency relation, and Sec.~\ref{sec:conclusion} summarizes the results.

\section{Scale-dependent collapse exterior}\label{sec:geometry}

\subsection{Metric, running coupling, and central curvature}

The exterior line element is
\begin{equation}
\dif s^2=-F(r)c^2\dif t^2+\frac{\dif r^2}{F(r)}+r^2\dif\Omega_2^2,
\label{eq:metric-physical}
\end{equation}
where $t$ is normalized by the asymptotic static clock and
$\dif\Omega_2^2=\dif\theta^2+\sin^2\theta\,\dif\phi^2$ is the
line element on the unit two-sphere.  The subscript distinguishes this
geometrical quantity from the scale-dependence parameter $\Omega$ below.
A convenient parametrization of the renormalization-group-improved running
Newton coupling is \cite{Bonanno:2000,Hassannejad:2024cbu}
\begin{equation}
G(r)=\frac{G_0r^3}{r^3+\Omega(r+\gamma\rs/2)},
\qquad
\rs=\frac{2G_0M}{c^2}.
\label{eq:running-G}
\end{equation}

Here $G_0=\lim_{r\to\infty}G(r)$ is the infrared Newton constant, $M$ is
the asymptotic gravitational mass, and $\Omega$ is the dimensionful
scale-dependence parameter, with dimensions of length squared.
The corresponding blackening function is
\begin{equation}
F(r)=1-\frac{2MG(r)}{c^2r}
=1-\frac{\rs r^2}{r^3+\Omega(r+\gamma\rs/2)}.
\label{eq:F-dimensional}
\end{equation}
We introduce the dimensionless variables

\begin{equation}
 x=\frac{r}{\rs},\qquad \wtil=\frac{\Omega}{\rs^2}.
\label{eq:dimensionless}
\end{equation}

The parameter $\gamma$ is not an additional running coupling; rather, it
specifies the interpolation used to identify the renormalization-group scale
through $k(r)=\xi/d(r)$.  In the proper-distance prescription introduced in
Ref.~\cite{Bonanno:2000}, the interpolating distance scale may be written as
\begin{equation}
d_{\rm int}(r)=
\left(\frac{r^3}{r+\gamma r_{\rm s}/2}\right)^{1/2}.
\label{eq:cutoff-distance}
\end{equation}
It satisfies $d_{\rm int}(r)\simeq r$ at large radius, whereas for
$r\ll r_{\rm s}$,
\begin{equation}
d_{\rm int}(r)=
\sqrt{\frac{2}{\gamma}}\,
\frac{r^{3/2}}{\sqrt{r_{\rm s}}}
\left[1+\Order\left(\frac{r}{r_{\rm s}}\right)\right].
\label{eq:cutoff-distance-small-r}
\end{equation}
On the other hand, the generalized radial proper distance evaluated in the
classical Schwarzschild geometry behaves as
\begin{equation}
d_{\rm Sch}(r)
=\int_0^r\frac{\dif r'}{\sqrt{|1-r_{\rm s}/r'|}}
=\frac{2}{3}\frac{r^{3/2}}{\sqrt{r_{\rm s}}}
\left[1+\Order\left(\frac{r}{r_{\rm s}}\right)\right].
\label{eq:schwarzschild-distance-small-r}
\end{equation}
The same leading coefficient follows from the remaining proper time of a
marginally bound radial geodesic approaching the center.  Matching the
small-radius coefficients therefore gives
\begin{equation}
\sqrt{\frac{2}{\gamma}}=\frac{2}{3},
\qquad\Longrightarrow\qquad
\gamma=\frac{9}{2}.
\label{eq:gamma-choice}
\end{equation}
Accordingly, we retain $\gamma>0$ in the analytic derivations and use
$\gamma=9/2$ in the numerical figures as the standard
proper-distance-matched benchmark.  This value should not be interpreted as a
universal prediction of asymptotic safety: it depends on the cutoff
identification, while the qualitative horizon structure is robust throughout
the $\gamma>0$ sector.  The cases $\gamma=0$ and $\gamma<0$ have different
central and pole structures and lie outside the classification adopted below.
With this domain understood, we write
\begin{equation}
F(x)=1-\frac{x^2}{\D(x)},
\qquad
\D(x)=x^3+\wtil\left(x+\frac{\gamma}{2}\right).
\label{eq:F-D}
\end{equation}
The physical lapse of the static exterior is
\begin{equation}
\N(x)=\sqrt{F(x)}.
\label{eq:lapse}
\end{equation}
At large radius,
\begin{equation}
F(x)=1-\frac1x+\frac{\wtil}{x^3}+\frac{\gamma\wtil}{2x^4}
-\frac{\wtil^2}{x^5}+\Order(x^{-6}),
\label{eq:F-weak}
\end{equation}
so the leading running correction appears at order $r^{-3}$.

\begin{figure}[t]
\centering
\includegraphics[width=\columnwidth]{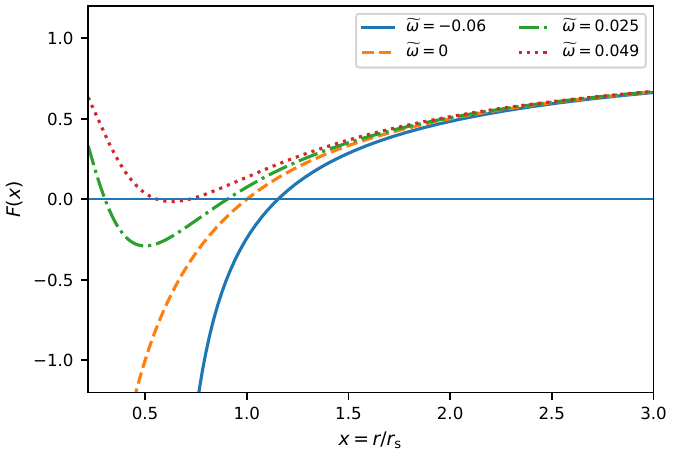}
\caption{Blackening factor for $\gamma=9/2$.  Zeros of $F$ are horizons.  For $\wtil=-0.06$, the curve diverges at $x_s=0.551910$ and has a shielded outer horizon at $x_+=1.153481$.  The Schwarzschild curve ($\wtil=0$) has $x_+=1$.  The positive-running branches have two horizons: $(x_-,x_+)=(0.302457,0.903405)$ for $\wtil=0.025$ and $(0.557768,0.717659)$ for $\wtil=0.049$.  Thus every curve in the legend is explicitly matched to its geometric structure.}
\label{fig:blackening}
\end{figure}

For $\gamma>0$ and $\wtil>0$, the center is curvature-finite: the independent orthonormal-frame components of the Riemann tensor, and hence the polynomial curvature invariants, remain finite.  Expanding Eq.~\eqref{eq:F-D} at $x=0$ gives
\begin{equation}
F(x)=1-\frac{2x^2}{\gamma\wtil}+\Order(x^3).
\label{eq:core-expansion}
\end{equation}
For a metric of the form \eqref{eq:metric-physical}, the Ricci scalar and the Kretschmann invariant are given by
\begin{align}
R&=-F_{,rr}-\frac{4F_{,r}}{r}+\frac{2(1-F)}{r^2},\label{eq:ricci-scalar-general}\\
K&=F_{,rr}^2+4\left(\frac{F_{,r}}{r}\right)^2
+4\left(\frac{1-F}{r^2}\right)^2.\label{eq:kretschmann-general}
\end{align}
Consequently,
\begin{equation}
R(0)=\frac{24}{\gamma\wtil\rs^2},
\qquad
K(0)=\frac{96}{\gamma^2\wtil^2\rs^4},
\label{eq:core-invariants}
\end{equation}
which are finite for fixed $\gamma>0$ and $\wtil>0$.  The first subleading corrections in the small-$x$ expansions of these two curvature invariants are
\begin{align}
R\rs^2&=\frac{24}{\gamma\wtil}
-\frac{80}{\gamma^2\wtil}x+\Order(x^2),\label{eq:core-R-next}\\
K\rs^4&=\frac{96}{\gamma^2\wtil^2}
-\frac{640}{\gamma^3\wtil^2}x+\Order(x^2).\label{eq:core-K-next}
\end{align}
The nonzero linear terms in Eqs.~\eqref{eq:core-R-next} and \eqref{eq:core-K-next} imply radial profiles of the form $R(r)=R_0+R_1r+\cdots$ and $K(r)=K_0+K_1r+\cdots$.  Since $r=\sqrt{X^2+Y^2+Z^2}$, such terms are not differentiable as Cartesian scalars at the origin.  Thus curvature finiteness does not by itself establish an analytic or $C^\infty$ Cartesian extension, nor does it prove geodesic completeness.  We therefore use ``curvature-finite core'' rather than ``regular center'' in the strong smoothness sense.  Negative $\wtil$ instead produces a finite-radius pole of $F$.  Figure~\ref{fig:blackening} illustrates the behavior of the lapse function \(F(x)\), for several values of \(\wtil\); its positive roots identify the corresponding horizon locations.

\subsection{Horizons, singular shell, and extremality}

Since $F=(\D-x^2)/\D$, horizons are positive roots of
\begin{equation}
\Pp(x)=x^3-x^2+\wtil x+\frac{\gamma\wtil}{2}=0
\label{eq:horizon-poly}
\end{equation}
for which $\D\neq0$.  A singular shell is a positive root of
\begin{equation}
\D(x_s)=x_s^3+\wtil\left(x_s+\frac{\gamma}{2}\right)=0.
\label{eq:singular-shell}
\end{equation}
For $\wtil<0$, $\D(0)<0$ and $\D(\infty)>0$, so a positive $x_s$ exists.  At that point $\Pp(x_s)=-x_s^2<0$, while $\Pp(\infty)>0$; hence the unique outer horizon obeys $x_+>x_s$ and shields the shell.  For $\wtil>0$, $\D(x)>0$ on $x>0$.

An extremal horizon is a double root,
\begin{equation}
\Pp(x_e)=0,
\qquad
\Pp'(x_e)=3x_e^2-2x_e+\wtil_e=0.
\label{eq:extremal-conditions}
\end{equation}
Eliminating $\wtil_e$ gives
\begin{equation}
2x_e^2-\left(1-\frac{3\gamma}{2}\right)x_e-\gamma=0,
\label{eq:xe-quadratic}
\end{equation}
so that
\begin{align}
 x_e&=\frac{1-3\gamma/2+\sqrt{(1-3\gamma/2)^2+8\gamma}}{4},
\label{eq:xe}\\
 \wtil_e&=x_e(2-3x_e).
\label{eq:we}
\end{align}
For $\gamma=9/2$,
\begin{equation}
 x_e=0.6400962673\ldots,
\qquad
 \wtil_e=0.0510228403\ldots.
\label{eq:critical-numerical}
\end{equation}
Thus $0<\wtil<\wtil_e$ has two positive horizons, $\wtil=\wtil_e$ is extremal, and $\wtil>\wtil_e$ is horizonless.

\begin{figure}[t]
\centering
\includegraphics[width=\columnwidth]{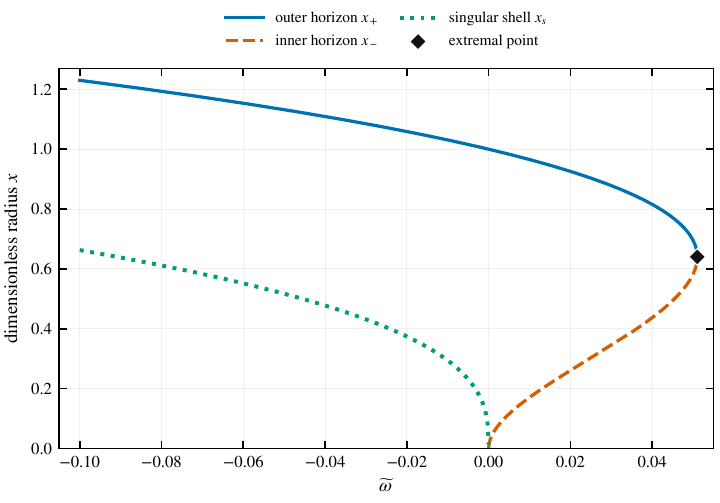}
\caption{Outer horizon (solid), inner horizon (dashed), and positive singular-shell radius (dotted) for $\gamma=9/2$.  The positive-running horizons merge at $(\wtil_e,x_e)$.  The negative-running shell remains inside the outer horizon.  The limiting endpoints $x_-\to0$ and $x_s\to0$ as $\wtil\to0^\pm$ are not admissible roots at $\wtil=0$.}
\label{fig:horizon-map}
\end{figure}

The complete horizon and shell map is shown in Fig.~\ref{fig:horizon-map}.  The negative-running branch contains one outer horizon outside the singular shell, whereas the positive-running branch contains inner and outer horizons that merge at $(\wtil_e,x_e)$.  As $\wtil\to0^+$, the inner radius tends to zero, and as $\wtil\to0^-$ the shell radius does likewise; neither limiting point represents an admissible root at $\wtil=0$, because $\D(0)=0$.

\begin{figure}[t]
\centering
\includegraphics[width=\columnwidth]{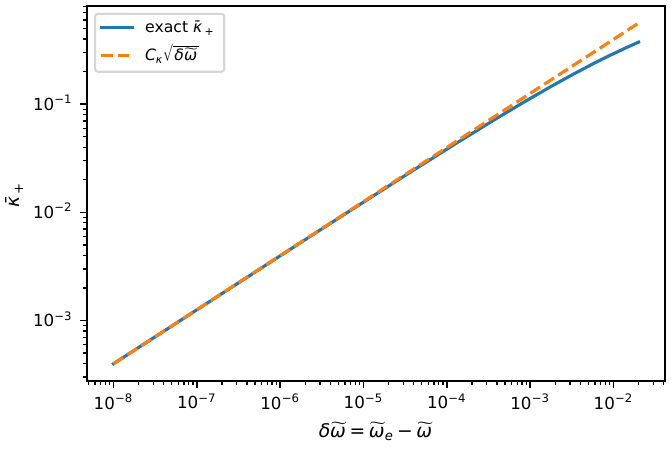}
\caption{Exact outer-horizon surface gravity and the square-root asymptotic law \eqref{eq:kappa-sqrt} for $\gamma=9/2$.  The relative error is below $1\%$ for $\delta\wtil\lesssim5\times10^{-6}$, is approximately $5\%$ at $\delta\wtil=2\times10^{-4}$, and grows outside the asymptotic regime.  The broad logarithmic range therefore displays both the near-extremal agreement and its breakdown.}
\label{fig:kappa}
\end{figure}

The surface gravity of a simple horizon is
\begin{equation}
\kappa_h=\frac{c^2}{2\rs}|F_x(x_h)|,
\qquad
\kb_h\equiv\frac{\kappa_h\rs}{c^2}
=\frac12\left|3-\frac{2}{x_h}+\frac{\wtil}{x_h^2}\right|.
\label{eq:surface-gravity}
\end{equation}
Writing $\delta\wtil=\wtil_e-\wtil>0$ and expanding about the double root gives
\begin{align}
 x_\pm&=x_e\pm
\sqrt{\frac{2(x_e+\gamma/2)}{6x_e-2}}\sqrt{\delta\wtil}
+\Order(\delta\wtil),
\label{eq:horizon-splitting}\\
\kb_+&=C_\kappa\sqrt{\delta\wtil}+\Order(\delta\wtil),
\label{eq:kappa-sqrt}
\end{align}
where
\begin{equation}
C_\kappa=\frac{\sqrt{2(x_e+\gamma/2)(6x_e-2)}}{2x_e^2}.
\label{eq:C-kappa}
\end{equation}
The square-root law is asymptotic rather than a global fit.  For $\gamma=9/2$, its relative error is approximately $5\%$ at $\delta\wtil=2\times10^{-4}$, falls below $1\%$ for $\delta\wtil\lesssim5\times10^{-6}$, and reaches approximately $50\%$ at $\delta\wtil=2\times10^{-2}$.  Figure~\ref{fig:kappa} displays both the genuine asymptotic agreement and its breakdown at larger $\delta\wtil$.

\subsection{Rindler region and extremal throat}

Let $\zeta=x-x_+$.  Near a nonextremal horizon,
\begin{equation}
F(x)=2\kb_+\zeta+\Order(\zeta^2),
\label{eq:Rindler-F}
\end{equation}
and the proper distance from the horizon is
\begin{equation}
\rho=\rs\int_{x_+}^{x}\frac{\dif x'}{\sqrt{F(x')}}
\simeq\rs\sqrt{\frac{2\zeta}{\kb_+}}.
\label{eq:proper-distance-nonext}
\end{equation}
Therefore,
\begin{equation}
\N\simeq\frac{\kappa_+\rho}{c^2}.
\label{eq:Rindler-lapse}
\end{equation}

At extremality, $F(x_e)=F_x(x_e)=0$ and
\begin{equation}
F_{xx}(x_e)=\frac{6x_e-2}{x_e^2}.
\label{eq:Fxx-ext}
\end{equation}
Hence
\begin{equation}
F(x)=\frac{(x-x_e)^2}{\ell_2^2}+\Order[(x-x_e)^3],
\qquad
\ell_2^2=\frac{x_e^2}{3x_e-1}.
\label{eq:extremal-F}
\end{equation}
With $y=r-r_e$ and $L_2=\rs\ell_2$, the leading metric is
\begin{equation}
\dif s^2\simeq-\frac{y^2}{L_2^2}c^2\dif t^2
+\frac{L_2^2}{y^2}\dif y^2+r_e^2\dif\Omega_2^2,
\label{eq:ads2s2}
\end{equation}
which is locally $\mathrm{AdS}_2\times S^2$ \cite{Kunduri:2013}.  For $\gamma=9/2$, $L_2/\rs=0.66724177\ldots$.

\begin{figure}[t]
\centering
\includegraphics[width=\columnwidth]{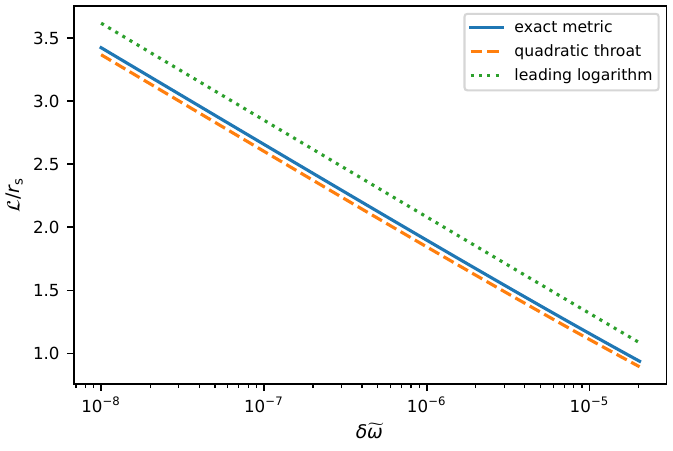}
\caption{Proper length of the throat segment $\zeta\in[\zeta_T,\zeta_{\rm UV}]$ for $\gamma=9/2$ and $\zeta_{\rm UV}=0.08$.  The exact-metric integral, the quadratic throat model, and its leading logarithm agree progressively as extremality is approached.}
\label{fig:throat}
\end{figure}

Near extremality, the throat is captured by
\begin{equation}
F(\zeta)\simeq\frac{\zeta(\zeta+\zeta_T)}{\ell_2^2},
\qquad
\zeta_T=2\kb_+\ell_2^2.
\label{eq:near-ext-throat}
\end{equation}
For $\zeta_T\ll\zeta_{\rm UV}\ll1$,
\begin{align}
\frac{\mathcal L_{\rm throat}}{\rs}
&=2\ell_2\left[
\operatorname{arsinh}\sqrt{\frac{\zeta}{\zeta_T}}
\right]_{\zeta_T}^{\zeta_{\rm UV}}
\label{eq:throat-length-exact}\\
&=\ell_2\ln\left(\frac{\zeta_{\rm UV}}{\zeta_T}\right)+\Order(\ell_2).
\label{eq:throat-length}
\end{align}
Because $\zeta_T\propto\sqrt{\wtil_e-\wtil}$, the proper throat length grows logarithmically.  Figure \ref{fig:throat} also compares the quadratic model with direct integration of the exact metric; only points satisfying $\zeta_T<0.2\zeta_{\rm UV}$ are shown.

The throat is a region of the physical four-dimensional exterior.  It modifies clock conversion, proper distance, and signal propagation.  Within the minimally coupled test-condensate model adopted here, and with the orthonormal-frame material parameters held fixed, it does not by itself renormalize the local Ginzburg--Landau coefficients.  Curvature-dependent effective couplings, for example $a_{\rm eff}=a+\chi_4R+\chi_3\,{}^{(3)}R+\cdots$, as well as strain- and tidal-induced changes of the material parameters, lie outside the present approximation.

\section{Covariant Josephson kinematics}\label{sec:covariant}

\subsection{Gauge-invariant phase and clock-referenced voltage}

Let $\Psi=|\Psi|\ee^{\imag\theta}$ and $q=2e>0$.  Here $\Psi$ is the complex superconducting order parameter, $|\Psi|$ is its modulus, $\theta$ is its condensate phase, $e>0$ is the elementary-charge magnitude, and $q=2e$ is the magnitude of the Cooper-pair charge.  The sign of the electromagnetic coupling is fixed by the convention in the following definition. We use the gauge-invariant condensate momentum \cite{Pantig:2025}
\begin{equation}
\Pi_\mu=\hbar\nabla_\mu\theta-qA_\mu.
\label{eq:cond-momentum}
\end{equation}
Following relativistic superfluid hydrodynamics and its charged-superconductor generalization \cite{Son:2000ht,Anandan:1984rte}, we adopt $u^\mu\Pi_\mu=\mu$ and $\Phi=-u^\mu A_\mu$, so that the local phase law is
\begin{equation}
\frac{\dif\theta}{\dif\tau}=\frac{\mu-q\Phi}{\hbar}.
\label{eq:phase-proper}
\end{equation}
Here $u^\mu$ is the condensate-frame four-velocity, $A_\mu$ is the electromagnetic four-potential, and our signed phase-chemical-potential convention is $\mu\equiv u^\mu\Pi_\mu$.  Reversing that convention reverses the definitions of $V^{\rm p}$ and $\varphi$ together and leaves all observable frequencies and currents unchanged.
A controlled electrochemical voltage therefore gives
\begin{equation}
\frac{\dif\theta}{\dif\tau}=\frac{q}{\hbar}V^{\rm p}.
\label{eq:local-ac}
\end{equation}

For two banks connected by a spacelike curve $\mathcal C$ oriented from bank 2 to bank 1,
\begin{equation}
\varphi=\theta_1-\theta_2-\frac{q}{\hbar}\int_{\mathcal C}A_\mu\dif x^\mu
\label{eq:gauge-phase}
\end{equation}
is gauge invariant.  If the connector carries no time-dependent flux \cite{Pantig:2025}, then
\begin{equation}
\frac{\dif\varphi}{\dif t}
=\frac{q}{\hbar}\left(z_1V_1^{\rm p}-z_2V_2^{\rm p}\right),
\qquad
z_i=\frac{\dif\tau_i}{\dif t}.
\label{eq:master-phase}
\end{equation}
We define the Killing-time-referenced control \cite{Pantig:2025}
\begin{equation}
V_i^{(t)}\equiv z_iV_i^{\rm p},
\label{eq:time-referenced-voltage}
\end{equation}
so that $\dot\varphi=q(V_1^{(t)}-V_2^{(t)})/\hbar$.  For a static terminal, $V^{(t)}$ is the usual redshifted voltage referred to infinity.  For a moving terminal it is a clock-conversion variable; the frequency received after electromagnetic propagation requires an additional Doppler/null-transfer factor derived in Sec. \ref{sec:collapse}.

\begin{figure*}[t]
\centering
\includegraphics[
  width=0.98\textwidth,
  trim=5 4 5 4,
  clip
]{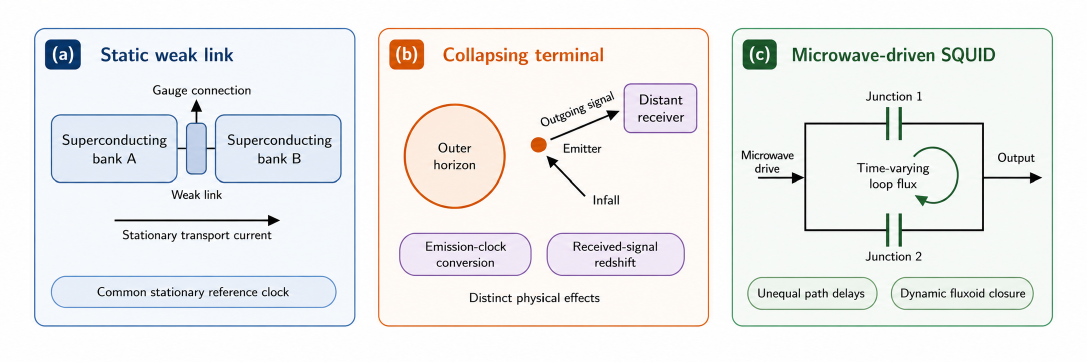}
\caption{Operational configurations considered in this work:
(a) a static weak link with a common stationary reference clock;
(b) a collapsing terminal, distinguishing emission-clock conversion
from the redshift of the signal received at infinity; and
(c) a microwave-driven two-junction dc SQUID with unequal propagation
delays and dynamical fluxoid closure.}
\label{fig:operational-josephson-schematic}
\end{figure*}

In static equilibrium, the Tolman--Klein condition generalized to a charged electrochemical potential gives \cite{Klein:1949te,Lima:2019brf,Anandan:1984rte}
\begin{equation}
\N(\mu-q\Phi)=\mathrm{const.}
\label{eq:TE-electrochemical}
\end{equation}
Equation~\eqref{eq:master-phase} describes a controlled nonequilibrium
bias.  The three operational configurations analyzed below are
summarized in Fig.~\ref{fig:operational-josephson-schematic}: a static
covariant weak link, a terminal comoving with the collapsing surface
and monitored through an outgoing signal, and a microwave-driven
two-junction SQUID subject to dynamical fluxoid closure.

\subsection{Static terminals}

The static weak-link configuration is represented schematically in
Fig.~\ref{fig:operational-josephson-schematic}(a).
\begin{equation}
U^\mu=\frac1\N(\partial_t)^\mu,
\qquad
\dif\tau=\N\dif t.
\label{eq:static-observer}
\end{equation}
For static terminals at $x_1$ and $x_2$, we define the signed Josephson
angular phase rate referred to the asymptotically normalized Killing time by
\begin{equation}
\begin{aligned}
\omega_{J,\infty}
&\equiv\frac{\dif\varphi}{\dif t}
=\frac{q}{\hbar}\left(V_1^{(t)}-V_2^{(t)}\right)\\
&=\frac{2e}{\hbar}
\left[\N(x_1)V_1^{\rm p}-\N(x_2)V_2^{\rm p}\right].
\end{aligned}
\label{eq:static-ac-law}
\end{equation}
The subscript $\infty$ specifies the clock normalization; it does not imply
an additional propagation transfer to a distant receiver.  The sign of
$\omega_{J,\infty}$ fixes the phase-winding direction, while the corresponding
nonnegative cyclic frequency is
$f_{J,\infty}=|\omega_{J,\infty}|/(2\pi)$.  If the voltages are already
calibrated against the asymptotic clock,
$V_i^{\rm p}=V_i^\infty/\N_i$, the flat-space conversion law is recovered.
Conversely, equal proper biases at separated terminals produce
\begin{equation}
\omega_{J,\infty}=\frac{2eV_0}{\hbar}[\N(x_1)-\N(x_2)].
\label{eq:equal-proper-bias}
\end{equation}
For a single compact junction, $x_1-x_2$ is microscopic and the lapse is effectively common; Eq. \eqref{eq:equal-proper-bias} is instead relevant to separated phase-coherent terminals or an extended interferometric circuit.

At a fixed exterior radius $x>x_+(\wtil)$, an expansion about Schwarzschild at fixed $x>1$ is valid provided
\begin{equation}
\epsilon_{\wtil}(x)\equiv
\left|\frac{\wtil}{F_{\rm Schw}(x)}
\left(\frac1{x^3}+\frac{\gamma}{2x^4}\right)\right|\ll1.
\label{eq:weak-running-validity}
\end{equation}
Under this condition,
\begin{equation}
\N_{\rm SD}=\N_{\rm Schw}\left[1+
\frac{\wtil}{2F_{\rm Schw}}
\left(\frac1{x^3}+\frac{\gamma}{2x^4}\right)
+\Order(\wtil^2)\right],
\label{eq:lapse-perturbation}
\end{equation}
where $F_{\rm Schw}=1-1/x$.

\begin{figure}[t]
\centering
\includegraphics[width=\columnwidth]{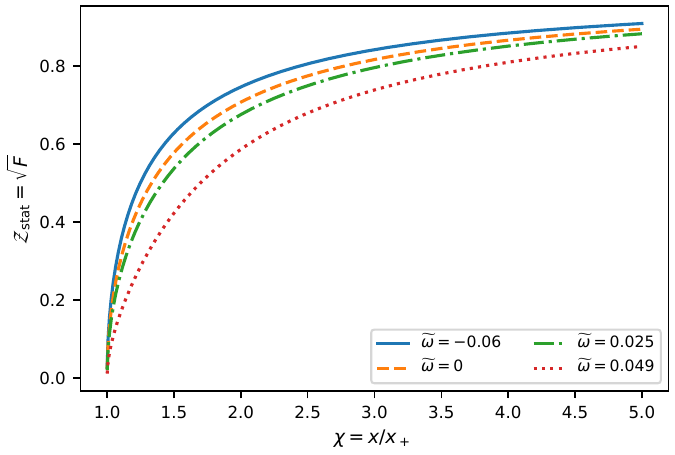}
\caption{Static one-lapse factor $\N=\sqrt F$ versus $X=x/x_+$ for $\gamma=9/2$ and the values of $\wtil$ shown in the legend.  At fixed locally calibrated proper bias or current, this factor maps the generated ac frequency or transported charge current to Killing time.  The comparison is made at equal $X$, not equal areal radius.}
\label{fig:static-redshift}
\end{figure}

Figure~\ref{fig:static-redshift} compares the one-lapse factor at equal normalized radius $X=x/x_+$ for several branches.  This normalization holds the radius relative to each branch's own outer horizon fixed and therefore should not be confused with the fixed-$x$ expansion in Eq.~\eqref{eq:lapse-perturbation}.  The plotted factor applies to a locally calibrated proper voltage or proper current; a control already calibrated against the asymptotic clock follows the distinct protocol described above.

\subsection{Conserved charge flux and power}

Let $j^\mu$ be a conserved charge current, $\nabla_\mu j^\mu=0$.  On a static slice with zero shift,
\begin{equation}
j^\mu=\rho U^\mu+\mathcal J^\mu,
\qquad U_\mu\mathcal J^\mu=0,
\label{eq:current-split}
\end{equation}
and
\begin{equation}
\partial_t(\sqrt\gamma\,\rho)
+\partial_i(\sqrt\gamma\,\N\mathcal J^i)=0.
\label{eq:3plus1-continuity}
\end{equation}
The charge crossing a two-surface $S$ per unit Killing time is
\begin{equation}
I_\infty[S]=\int_S\N\mathcal J^is_i\dif A,
\label{eq:I-infinity}
\end{equation}
whereas the local proper-time current is
\begin{equation}
I_{\rm p}[S]=\int_S\mathcal J^is_i\dif A.
\label{eq:I-proper}
\end{equation}
For a compact static junction, the emitted power at infinity $P_\infty$ 
\begin{equation}
I_\infty=\N_JI_{\rm p},
\qquad
P_\infty=V_\infty I_\infty=\N_J^2P_{\rm p}=F_JP_{\rm p}.
\label{eq:current-power-map}
\end{equation}
One lapse converts local energy per charge and the second converts proper time into Killing time.

\section{Microwave locking and null propagation}\label{sec:ac}

\subsection{Voltage-biased harmonic locking}

Consider a differential drive referred to the asymptotic clock,
\begin{equation}
V_1^\infty(t)-V_2^\infty(t)=V_{\rm dc}^\infty+
\Re\left(\mathcal V_{\rm mw}^\infty\ee^{-\imag\Omega_\infty t}\right).
\label{eq:rf-drive}
\end{equation}

Here $\Omega_\infty$ is the angular frequency of the applied
microwave drive measured with respect to the asymptotic Killing time
$t$; it is unrelated to the dimensionful gravitational parameter $\Omega$.
Writing $\mathcal V_{\rm mw}^{\infty}
=|\mathcal V_{\rm mw}^{\infty}|\ee^{\imag\psi}$ fixes the phase convention,
so the real microwave drive is
$|\mathcal V_{\rm mw}^{\infty}|\cos(\Omega_\infty t-\psi)$.
The phase is
\begin{equation}
\varphi(t)=\varphi_0+\omega_0t+a\sin(\Omega_\infty t-\psi),
\label{eq:rf-phase}
\end{equation}
with
\begin{equation}
\omega_0=\frac{2eV_{\rm dc}^\infty}{\hbar},
\qquad
a=\frac{2e|\mathcal V_{\rm mw}^\infty|}{\hbar\Omega_\infty}.
\label{eq:omega-a}
\end{equation}
A zero-frequency harmonic occurs at
\begin{equation}
\omega_0=n\Omega_\infty,
\qquad
V_{\rm dc}^\infty=n\frac{\hbar\Omega_\infty}{2e}.
\label{eq:step-condition}
\end{equation}
With the phase convention in Eq. \eqref{eq:rf-phase}, the time-averaged Josephson current is
\begin{equation}
\langle I\rangle_n=(-1)^n I_cJ_n(a)
\sin(\varphi_0+n\psi).
\label{eq:step-height}
\end{equation}
The factor $(-1)^n$ may be absorbed into a redefinition of $\varphi_0$, but the sign of the propagation phase is fixed by the convention in Eq. \eqref{eq:rf-phase}.  Equation \eqref{eq:step-condition} identifies the voltage-biased locking harmonics.  The widths and stability of current-biased Shapiro plateaus require the RCSJ dynamics derived in Sec. \ref{sec:rcsj}.

When a source is imposed locally, its asymptotic amplitude and frequency must both be converted using the source clock.  A locally calibrated drive and a drive already referenced to infinity are distinct protocols.

\subsection{Exact radial-null propagation kernel in the eikonal limit}

For an outgoing radial null ray with $x_b>x_a>x_+$,
\begin{equation}
\frac{\dif t}{\dif x}=\pm\frac{\rs}{cF(x)}.
\label{eq:null-time}
\end{equation}
Since $\D=\Pp+x^2$,
\begin{equation}
\frac1F=1+\frac{x^2}{\Pp(x)}.
\label{eq:one-over-F}
\end{equation}
For three distinct roots $x_j$ of $\Pp$,
\begin{equation}
\frac{x^2}{\Pp(x)}=\sum_{j=1}^3\frac{A_j}{x-x_j},
\qquad
A_j=\frac{x_j^2}{\Pp'(x_j)}.
\label{eq:Aj}
\end{equation}
Thus
\begin{equation}
\Delta t_{ab}=\frac{\rs}{c}\left[x_b-x_a+
\sum_{j=1}^3A_j\ln\left(\frac{x_b-x_j}{x_a-x_j}\right)\right],
\label{eq:exact-propagation}
\end{equation}
where conjugate terms are combined to give a real result.  In the Schwarzschild limit, where the cubic has a repeated zero root, the regular limiting expression is
\begin{equation}
\Delta t_{ab}^{\rm Schw}=\frac{\rs}{c}
\left[x_b-x_a+\ln\left(\frac{x_b-1}{x_a-1}\right)\right].
\label{eq:schw-propagation}
\end{equation}
Near a simple outer horizon,
\begin{equation}
A_+=\frac1{F_x(x_+)}=\frac1{2\kb_+},
\qquad
\Delta t\sim-\frac{\rs}{2c\kb_+}\ln(x_a-x_+).
\label{eq:near-horizon-time}
\end{equation}
Equation~\eqref{eq:near-horizon-time} is a local Rindler result.  Its formal $\kb_+\to0$ limit is nonuniform because the Rindler cap shrinks with the throat scale $\zeta_T=2\kb_+\ell_2^2$.  Using Eq.~\eqref{eq:near-ext-throat}, the dimensionless travel time across the near-extremal throat from $\zeta=\delta$ to $\zeta=\zeta_{\rm UV}$ is
\begin{equation}
\widehat{\Delta t}_{\rm th}
\equiv\frac{c\Delta t_{\rm th}}{\rs}
=
\frac{\ell_2^2}{\zeta_T}
\ln\!\left[
\frac{\zeta_{\rm UV}(\delta+\zeta_T)}
{\delta(\zeta_{\rm UV}+\zeta_T)}
\right].
\label{eq:uniform-throat-travel-time}
\end{equation}
For two rays emitted at fixed offsets $0<\delta_1<\delta_2$ and received at the same outer endpoint, their throat contribution satisfies
\begin{equation}
\Delta\widehat t_{12}
\equiv\frac{c(\Delta t_1-\Delta t_2)}{\rs}
=
\frac{\ell_2^2}{\zeta_T}
\ln\!\left[
\frac{\delta_2(\delta_1+\zeta_T)}
{\delta_1(\delta_2+\zeta_T)}
\right].
\label{eq:uniform-differential-delay}
\end{equation}
The two limiting regimes are
\begin{equation}
\Delta\widehat t_{12}\simeq
\begin{cases}
\displaystyle
\frac{1}{2\kb_+}\ln\!\left(\frac{\delta_2}{\delta_1}\right),
& \delta_i\ll\zeta_T,\\[2ex]
\displaystyle
\ell_2^2\left(\frac1{\delta_1}-\frac1{\delta_2}\right),
& \zeta_T\ll\delta_1<\delta_2.
\end{cases}
\label{eq:differential-delay-regimes}
\end{equation}
Thus the local $1/(2\kb_+)$ enhancement applies only when the emission points remain inside the shrinking Rindler cap.  At fixed nonzero coordinate offsets, the limit $\kb_+\to0$ instead crosses over to the pole-controlled extremal result; the limits $\delta_i\to0$ and $\kb_+\to0$ do not commute.  At exact extremality, the tortoise coordinate contains a pole proportional to $-(x-x_e)^{-1}$ together with logarithmic terms; its closed form is given in Appendix~\ref{app:propagation}.

For fixed endpoints $x_b>x_a>\max\{1,x_+\}$ away from the shifted horizon, the weak-running correction is
\begin{equation}
\Delta t_{ab}=\Delta t_{ab}^{\rm Schw}
-\frac{\rs}{c}\wtil[\mathcal K(x_b)-\mathcal K(x_a)]
+\Order(\wtil^2),
\label{eq:weak-w-time}
\end{equation}
where
\begin{equation}
\mathcal K(x)=(\gamma+1)\ln\left(\frac{x}{x-1}\right)
-\frac{\gamma}{2x}-\frac{\gamma+2}{2(x-1)}.
\label{eq:K-function}
\end{equation}
The expansion is nonuniform at $x=1$.

For two microwave feed paths,
\begin{equation}
\Delta\psi=\Omega_\infty(\Delta t_1-\Delta t_2).
\label{eq:differential-phase}
\end{equation}
This phase modifies locked-current amplitudes and two-junction interference, but not the voltage quantization in Eq.~\eqref{eq:step-condition}.  Equation~\eqref{eq:differential-phase} is an eikonal transfer law: the null travel time is exact for radial rays, whereas its direct use as a finite-frequency microwave phase requires
\begin{equation}
\epsilon_{\rm GO}(x)
\equiv
\frac{c}{\Omega_{\rm loc}(x)L_{\rm curv}(x)}\ll1,
\qquad
\Omega_{\rm loc}(x)=\frac{\Omega_\infty}{\N(x)},
\label{eq:geometric-optics-validity}
\end{equation}
throughout the relevant path.  Here $L_{\rm curv}$ denotes the shortest local curvature or background-inhomogeneity scale.  A narrow signal bandwidth and negligible dispersive response of the feed network are also assumed.  Subleading Maxwell-wave effects can include curvature scattering, diffraction, wave tails, and polarization-dependent transport.  Accordingly, Eq.~\eqref{eq:exact-propagation} is exact as a radial-null travel-time kernel, but not as a complete finite-frequency Maxwell transfer function.

\begin{figure}[t]
\centering
\includegraphics[width=\columnwidth]{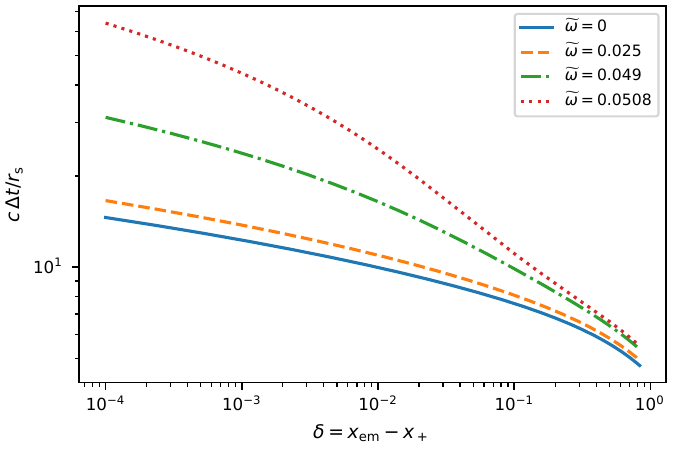}
\caption{Dimensionless outgoing radial-null coordinate travel time $c\Delta t/r_s$ from $x_{\rm em}=x_++\delta$ to $x=5$, for $\gamma=9/2$ and the values of $\wtil$ shown in the legend.  Over the displayed nonextremal range, the near-horizon logarithmic coefficient $1/(2\kb_+)$ produces the enhanced delay.  At fixed nonzero $\delta$, the arbitrarily near-extremal limit instead crosses over to the pole-controlled extremal kernel.}
\label{fig:propagation}
\end{figure}

The exact outgoing coordinate travel times are displayed in Fig.~\ref{fig:propagation}.  At $\delta=10^{-4}$, one finds $c\Delta t/r_s=14.5965$ for Schwarzschild and $c\Delta t/r_s=64.0202$ for $\wtil=0.0508$.  Over the displayed nonextremal range, the steep near-horizon growth is governed by the local logarithmic coefficient $1/(2\kb_+)$.  At fixed nonzero $\delta$, however, a still closer approach to extremality eventually leaves the shrinking Rindler cap and crosses over to the pole-controlled limit in Eqs.~\eqref{eq:uniform-throat-travel-time}--\eqref{eq:differential-delay-regimes}.  The curves converge at larger emission radii because the near-horizon segment contributes less of the total delay.

\section{Collapsing terminal: coordinate and received signals}\label{sec:collapse}

The distinction between emission-clock conversion and the redshift of
the received outgoing signal is illustrated in
Fig.~\ref{fig:operational-josephson-schematic}(b). In the marginally bound specialization of the Oppenheimer--Snyder-like collapse model, the stellar surface follows a radial geodesic with specific Killing energy $\mathcal E=1$ \cite{Hassannejad:2024cbu}; the original Oppenheimer--Snyder construction provides the historical collapse framework \cite{Oppenheimer:1939}.  More generally,
\begin{equation}
\frac{\dif t}{\dif\tau}=\frac{\mathcal E}{F},
\qquad
\frac1{c^2}\left(\frac{\dif R}{\dif\tau}\right)^2=\mathcal E^2-F.
\label{eq:surface-geodesic}
\end{equation}
For $X=R/\rs$ and $\widetilde\tau=c\tau/\rs$, the marginally bound branch obeys
\begin{equation}
\frac{\dif X}{\dif\widetilde\tau}=-\sqrt{1-F(X)}
=-\frac{X}{\sqrt{\D(X)}}.
\label{eq:collapse-equation}
\end{equation}

The conversion from proper time to the exterior coordinate $t$ is
\begin{equation}
z_t\equiv\frac{\dif\tau}{\dif t}=\frac{F}{\mathcal E}.
\label{eq:surface-z-coordinate}
\end{equation}
Consequently, a local differential voltage produces the coordinate-time phase rate
\begin{equation}
\left.\frac{\dif\varphi}{\dif t}\right|_{\rm em}
=\frac{2e}{\hbar}\frac{F(X)}{\mathcal E}\Delta V^{\rm p}.
\label{eq:surface-josephson-coordinate}
\end{equation}
For $\mathcal E=1$, this is the $F$ factor obtained by comparing the emitter's clock with the Killing coordinate.  It is not yet the frequency of an outgoing electromagnetic signal received at infinity.

Let
\begin{equation}
u=t-\frac{r_*}{c},
\qquad
r_*=\rs\int^x\frac{\dif x'}{F(x')}
\label{eq:retarded-time}
\end{equation}
be retarded time.  Along an infalling radial geodesic,
\begin{equation}
\frac{\dif u}{\dif\tau}
=\frac{\mathcal E+\sqrt{\mathcal E^2-F}}{F},
\label{eq:du-dtau}
\end{equation}
so the received clock factor is
\begin{equation}
z_{\rm out}\equiv\frac{\dif\tau}{\dif u}
=\frac{F}{\mathcal E+\sqrt{\mathcal E^2-F}}
=\mathcal E-\sqrt{\mathcal E^2-F}.
\label{eq:received-redshift}
\end{equation}
For a marginally bound surface,
\begin{equation}
z_{\rm out}=1-\sqrt{1-F},
\qquad
\omega_{J,\infty}^{\rm rec}
\equiv\frac{\dif\varphi}{\dif u}
=\frac{2e}{\hbar}z_{\rm out}\Delta V^{\rm p}.
\label{eq:received-josephson}
\end{equation}
Near a simple horizon, $z_{\rm out}=F/(2\mathcal E)+\Order(F^2)$: the $F$ scaling is robust, but the normalization differs from the coordinate-time expression.

Continuing with the marginally bound trajectory ($\mathcal E=1$), for a nonextremal horizon, $X-X_+\simeq\widetilde\tau_h-\widetilde\tau$ and
\begin{equation}
F\simeq2\kb_+(\widetilde\tau_h-\widetilde\tau).
\label{eq:F-propertime-nonext}
\end{equation}
The coordinate-time phase rate decays as
\begin{equation}
F(t)\propto\exp(-2\kappa_+t/c),
\label{eq:F-Killing-nonext}
\end{equation}
whereas retarded time satisfies $\widetilde u\equiv cu/\rs\simeq-\kb_+^{-1}\ln(\widetilde\tau_h-\widetilde\tau)$.  The received signal therefore obeys
\begin{equation}
z_{\rm out}(u)\propto\exp(-\kappa_+u/c).
\label{eq:received-nonext}
\end{equation}
At exact extremality,
\begin{equation}
F\simeq\frac{(\widetilde\tau_h-\widetilde\tau)^2}{\ell_2^2},
\qquad
\widetilde u\simeq\frac{2\ell_2^2}{\widetilde\tau_h-\widetilde\tau},
\label{eq:extremal-retarded}
\end{equation}
which gives
\begin{equation}
z_{\rm out}(u)\simeq\frac{2\ell_2^2}{\widetilde u^2}
\propto u^{-2}.
\label{eq:received-extremal}
\end{equation}
The transition from exponential to algebraic decay is a geometric consequence of the double horizon.

\begin{figure}[t]
\centering
\includegraphics[width=\columnwidth]{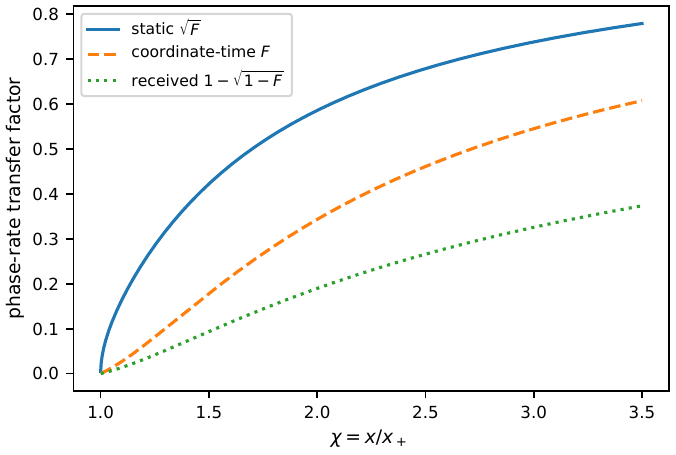}
\caption{Transfer factors for $\gamma=9/2$, $\wtil=0.049$, and a marginally bound infalling source, $\mathcal E=1$.  The static lapse is $\sqrt F$, the emitter's proper-to-coordinate-time factor is $F$, and the redshift of an outgoing signal received at infinity is $1-\sqrt{1-F}$.  The last two are both $\Order(F)$ near the horizon but differ by a finite Doppler normalization.}
\label{fig:transfer-factors}
\end{figure}

Figure~\ref{fig:transfer-factors} compares the three transfer factors for the representative branch $\gamma=9/2$, $\wtil=0.049$, and a marginally bound infalling source with $\mathcal E=1$.  The static factor behaves as $\sqrt F$, whereas both the coordinate-time factor and the received-signal factor are of order $F$ near the horizon.  Their leading ratio is finite, $z_{\rm out}/F\to1/2$, and represents the additional outgoing Doppler normalization.

The result assumes radial outgoing propagation, an electrostatic local bias, and negligible time-dependent magnetic flux through the moving connector.  A nonradial link or a nonzero magnetic field introduces additional propagation and motional-emf terms.

\section{Weak-link and RCSJ dynamics}\label{sec:dc}

\subsection{Local covariant Ginzburg--Landau problem}

On a static slice with spatial metric $\gamma_{ij}$, define
\begin{equation}
\mathcal D_i=D_i-\frac{\imag q}{\hbar}A_i.
\label{eq:gauge-spatial-derivative}
\end{equation}
We adopt the minimally coupled test-condensate functional, with $a$, $b$, and $m_*$ defined in the local orthonormal material frame and treated as prescribed material parameters.  Explicit curvature couplings, strain response, and tidal renormalization of these coefficients are omitted.  The local proper free energy is \cite{Ginzburg:1950sr,Gennes_1966,tinkham1975introduction}
\begin{multline}
\mathcal F_{\rm p}=\int\dif^3x\sqrt\gamma\Bigg[
 a|\Psi|^2+\frac b2|\Psi|^4\\
+\frac{\hbar^2}{2m_*}\gamma^{ij}
(\mathcal D_i\Psi)^*(\mathcal D_j\Psi)\Bigg].
\label{eq:GL-functional}
\end{multline}
The Killing-energy functional contains an additional lapse.  Across a short junction, that lapse is constant to leading order provided
\begin{equation}
L\ll\ell_\N,
\qquad
\ell_\N\equiv\frac{\N}{|\bm D\N|}
=\frac{2\rs\sqrt F}{|F_x|}.
\label{eq:lapse-length}
\end{equation}
Varying Eq. \eqref{eq:GL-functional} gives
\begin{equation}
-\frac{\hbar^2}{2m_*\sqrt\gamma}\mathcal D_i
(\sqrt\gamma\,\gamma^{ij}\mathcal D_j\Psi)
+a\Psi+b|\Psi|^2\Psi=0,
\label{eq:GL-equation}
\end{equation}
with proper current density
\begin{equation}
\mathcal J^i=\frac{q\hbar}{m_*}
\Im(\Psi^*\gamma^{ij}\mathcal D_j\Psi).
\label{eq:GL-current}
\end{equation}

Let $s$ be proper Gaussian-normal distance across a normal barrier of thickness $L$, and write the spatial metric near the interfaces as
\begin{equation}
\dif\ell^2=\dif s^2+h_{AB}(s,y)\dif y^A\dif y^B.
\label{eq:barrier-gaussian-metric}
\end{equation}
With $h_\perp=\det h_{AB}$, the normal part of the scalar Laplacian is

\begin{equation}
\begin{aligned}
\frac{1}{\sqrt{h_\perp}}\frac{\dif}{\dif s}
\left(\sqrt{h_\perp}\frac{\dif\Psi}{\dif s}\right)
&=\frac{\dif^2\Psi}{\dif s^2}
+\mathcal H_s\frac{\dif\Psi}{\dif s},\\
\mathcal H_s&\equiv\frac{\dif}{\dif s}\ln\sqrt{h_\perp}.
\end{aligned}
\label{eq:barrier-mean-curvature}
\end{equation}

Thus a generic curved interface produces a first-order correction controlled by the proper surface-curvature radius $R_{\rm surf}\equiv|\mathcal H_s|^{-1}$.  In a local gauge with $A_s=0$, and for a locally planar short barrier satisfying $L/\ell_\N\ll1$, $L/R_{\rm surf}\ll1$, and $L^2/R_c^2\ll1$, the linearized equation reduces to
\begin{equation}
-\xi_N^2\frac{\dif^2\Psi}{\dif s^2}+\Psi=0,
\qquad
\xi_N^2=\frac{\hbar^2}{2m_*a_N},
\label{eq:linear-GL}
\end{equation}
up to $\Order(L/\ell_\N,L/R_{\rm surf},L^2/R_c^2)$.  For interface values $\Psi_L$ and $\Psi_R$,
\begin{equation}
\Psi(s)=\Psi_L\frac{\sinh[(L-s)/\xi_N]}{\sinh(L/\xi_N)}
+\Psi_R\frac{\sinh(s/\xi_N)}{\sinh(L/\xi_N)}.
\label{eq:barrier-solution}
\end{equation}
Choosing the positive current orientation consistently with the phase definition gives
\begin{equation}
I_{\rm p}=I_c^{\rm p}\sin\varphi,
\qquad
I_c^{\rm p}=A_{\rm p}\frac{q\hbar}{m_*}
\frac{|\Psi_L||\Psi_R|}{\xi_N\sinh(L/\xi_N)}.
\label{eq:proper-CPR}
\end{equation}
The sign reverses if the barrier normal is reversed.  All lengths and areas are proper quantities.

The current at infinity is therefore
\begin{equation}
I_{c,\infty}=\N_JI_c^{\rm p}
\left[1+\Order\left(
\frac{L}{\ell_\N},
\frac{L}{R_{\rm surf}},
\frac{L^2}{R_c^2}\right)\right].
\label{eq:asymptotic-Ic}
\end{equation}
At fixed proper junction parameters,
\begin{equation}
\frac{I_{c,\infty}^{\rm SD}-I_{c,\infty}^{\rm Schw}}
{I_{c,\infty}^{\rm Schw}}
=\frac{\wtil}{2F_{\rm Schw}(x_J)}
\left(\frac1{x_J^3}+\frac{\gamma}{2x_J^4}\right)+\Order(\wtil^2).
\label{eq:Ic-fractional}
\end{equation}
At fixed proper junction parameters, and only outside any state-dependent thermal or mechanical exclusion region, the formal nonextremal near-horizon scaling is
\begin{equation}
I_{c,\infty}\simeq\frac{\kappa_+\rho}{c^2}I_c^{\rm p},
\qquad
P_\infty\simeq\left(\frac{\kappa_+\rho}{c^2}\right)^2P_{\rm p}.
\label{eq:near-horizon-current-power}
\end{equation}
In Hartle--Hawking equilibrium, $I_c^{\rm p}$ must instead be understood as the temperature-dependent quantity $I_c^{\rm p}[T_{\rm p}(\rho)]$ and vanishes when $T_{\rm p}\geq T_c$.  Therefore, Eq.~\eqref{eq:near-horizon-current-power} cannot be extrapolated through the superconducting exclusion layer derived in Sec.~\ref{sec:validity}.

\subsection{Redshifted RCSJ equation}\label{sec:rcsj}

For a static junction over which $\N_J$ is constant, the local RCSJ relation is \cite{Stewart:1968,McCumber:1968}
\begin{equation}
I_{\rm p}=I_c^{\rm p}\sin\varphi+\frac{V^{\rm p}}{R}
+C\frac{\dif V^{\rm p}}{\dif\tau}.
\label{eq:local-rcsj}
\end{equation}
Using $I_\infty=\N_JI_{\rm p}$, $V_\infty=\N_JV^{\rm p}$, and $\dif\tau=\N_J\dif t$, one obtains
\begin{equation}
I_\infty=\N_JI_c^{\rm p}\sin\varphi+\frac{V_\infty}{R}
+\frac{C}{\N_J}\frac{\dif V_\infty}{\dif t}.
\label{eq:asymptotic-rcsj-voltage}
\end{equation}
Since $V_\infty=(\hbar/q)\dot\varphi$,
\begin{equation}
\frac{\hbar C}{q\N_J}\ddot\varphi
+\frac{\hbar}{qR}\dot\varphi
+\N_JI_c^{\rm p}\sin\varphi=I_\infty.
\label{eq:asymptotic-rcsj}
\end{equation}
Thus the Killing-time parameters are
\begin{equation}
I_c^{(t)}=\N_JI_c^{\rm p},
\qquad
C^{(t)}=\frac{C}{\N_J},
\qquad
R^{(t)}=R.
\label{eq:effective-rcsj-parameters}
\end{equation}
The plasma frequency redshifts as
\begin{equation}
\omega_{p,\infty}=\N_J\sqrt{\frac{qI_c^{\rm p}}{\hbar C}}
=\N_J\omega_{p,\rm p},
\label{eq:plasma-redshift}
\end{equation}
while the Stewart--McCumber parameter
\begin{equation}
\beta_c
=\frac{qI_c^{\rm p}R^2C}{\hbar}
=\frac{qI_c^{(t)}[R^{(t)}]^2C^{(t)}}{\hbar}
\label{eq:beta-c}
\end{equation}
remains invariant because $I_c^{(t)}C^{(t)}=I_c^{\rm p}C$.  This invariance holds for a compact static element; a distributed circuit spanning an appreciable lapse gradient requires a spatially resolved model.

\section{Two-junction superconducting interference}\label{sec:squid}

\subsection{dc SQUID envelope}
\FloatBarrier

Consider a loop with two short junctions.  We neglect both the geometric self-inductance and the kinetic inductance of the superconducting banks, and we assume that bulk gauge-invariant phase gradients along the arms are negligible.  Under these assumptions, the complete fluxoid relation reduces to
\begin{equation}
\varphi_2-\varphi_1=2\pi\frac{\Phi_{\rm ext}}{\PhiQ}+2\pi m,
\qquad
\PhiQ=\frac{h}{2e}.
\label{eq:flux-constraint}
\end{equation}
Magnetic flux is a spatial holonomy and carries no lapse factor.  Define
\begin{equation}
\mathcal I_i=\N_iI_{ci}^{\rm p}.
\label{eq:arm-amplitude}
\end{equation}
Then
\begin{equation}
I_\infty=\mathcal I_1\sin\varphi_1+\mathcal I_2\sin\varphi_2,
\label{eq:squid-current}
\end{equation}
and maximization over the common phase gives
\begin{equation}
I_{c,\infty}^{\rm dc}(f)=
\sqrt{\mathcal I_1^2+\mathcal I_2^2+2\mathcal I_1\mathcal I_2\cos(2\pi f)},
\qquad
f=\frac{\Phi_{\rm ext}}{\PhiQ}.
\label{eq:dc-squid-envelope}
\end{equation}
The maxima remain at integer $f$ and the minima at half-integer $f$.  A lapse imbalance changes their amplitudes but does not translate the pattern in the zero-total-inductance model:
\begin{equation}
I_{\rm max}=\mathcal I_1+\mathcal I_2,
\qquad
I_{\rm min}=|\mathcal I_1-\mathcal I_2|.
\label{eq:squid-extrema}
\end{equation}
For identical proper junctions and $\N_{1,2}=\bar\N(1\mp\eta)$,
\begin{equation}
I_{c,\infty}^{\rm dc}=2\bar\N I_0
\sqrt{\cos^2(\pi f)+\eta^2\sin^2(\pi f)}.
\label{eq:squid-small-imbalance}
\end{equation}
Away from the nodes, the deformation begins at $\Order(\eta^2)$; the residual node current is linear in $|\eta|$.

\begin{figure}[t]
\centering
\includegraphics[width=\columnwidth]{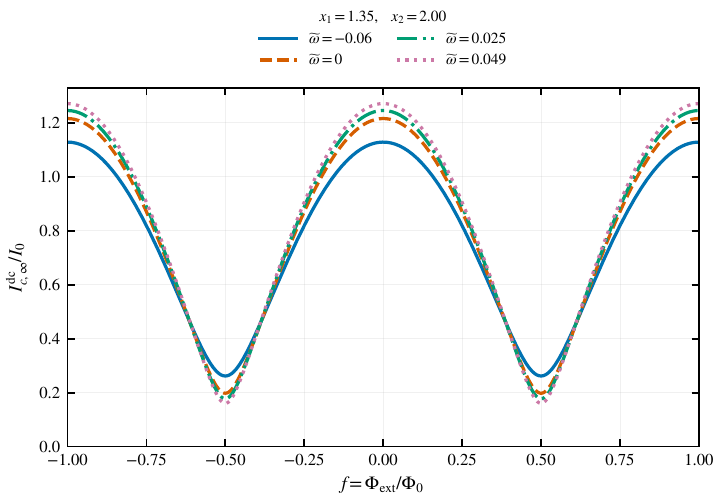}
\caption{dc SQUID envelope for $\gamma=9/2$ and identical proper junctions at $x_1=1.35$ and $x_2=2.00$, for the values of $\wtil$ shown in the legend.  The mean lapse changes the lobe amplitudes, while the radial lapse imbalance produces nonzero half-integer minima.  The integer maxima and half-integer minima remain at fixed flux positions in the zero-total-inductance limit.}
\label{fig:squid-dc}
\end{figure}

Figure~\ref{fig:squid-dc} evaluates Eq.~\eqref{eq:dc-squid-envelope} for identical proper junctions at $x_1=1.35$ and $x_2=2.00$.  For example, the Schwarzschild curve has $I_{\max}/I_0=1.2163$ and $I_{\min}/I_0=0.1979$, while the $\wtil=0.049$ curve has $I_{\max}/I_0=1.2717$ and $I_{\min}/I_0=0.1603$. Thus the mean lapse controls the overall lobe height, whereas the difference between the two lapse-weighted arm amplitudes produces the residual node current.

\subsection{Microwave-driven dc SQUID}

A conventional rf SQUID contains one junction in an inductive loop.  The two-junction device considered here is instead a dc SQUID under microwave drive. The microwave-driven two-junction configuration, including unequal
propagation paths and dynamical fluxoid closure, is illustrated in
Fig.~\ref{fig:operational-josephson-schematic}(c).  In time-dependent operation, fluxoid closure must hold instantaneously:
\begin{equation}
\varphi_2(t)-\varphi_1(t)
=2\pi\frac{\Phi_{\rm tot}(t)}{\PhiQ}+2\pi m.
\label{eq:dynamic-fluxoid}
\end{equation}
Consequently,
\begin{equation}
\dot\varphi_2-\dot\varphi_1
=\frac{2\pi}{\PhiQ}\dot\Phi_{\rm tot}
=\frac{q}{\hbar}\left[V_2^{(t)}(t)-V_1^{(t)}(t)\right],
\label{eq:dynamic-flux-balance}
\end{equation}
where $V_i^{(t)}$ is the Killing-time-referenced voltage across junction $i$.  Equation~\eqref{eq:dynamic-flux-balance} is the circuit form of Faraday closure and prevents the two junction phases and the loop flux from being prescribed independently.

Write
\begin{equation}
\Phi_{\rm tot}(t)=\Phi_{\rm dc}+\Phi_{\rm ac}(t),
\qquad
f\equiv\frac{\Phi_{\rm dc}}{\PhiQ}.
\label{eq:dc-ac-loop-flux}
\end{equation}
On the $n$th locking harmonic, let
$a_i\equiv q|\mathcal V_{{\rm mw},i}^{(t)}|/(\hbar\Omega_\infty)$
denote the dimensionless phase-modulation index of arm $i$, where
$\mathcal V_{{\rm mw},i}^{(t)}$ is its Killing-time-referenced microwave
amplitude.  We assume equal effective amplitudes, $a_1=a_2\equiv a$, and write
\begin{equation}
\varphi_i(t)=\varphi_{0i}+n\Omega_\infty t
+a\sin(\Omega_\infty t-\psi_i),
\qquad
\Delta\psi\equiv\psi_1-\psi_2.
\label{eq:two-arm-locked-phases}
\end{equation}
With $\varphi_{02}-\varphi_{01}=2\pi f+2\pi m$, the instantaneous constraint \eqref{eq:dynamic-fluxoid} requires
\begin{equation}
\frac{\Phi_{\rm ac}(t)}{\PhiQ}
=\frac{a}{2\pi}\left[
\sin(\Omega_\infty t-\psi_2)
-\sin(\Omega_\infty t-\psi_1)\right].
\label{eq:required-ac-loop-flux}
\end{equation}
Thus a differential propagation phase is consistent only when the microwave network supplies or accommodates the associated ac loop flux.  If $\Phi_{\rm tot}$ is required to be strictly static in the zero-total-inductance limit, equal nonzero modulation indices instead require $\psi_1=\psi_2$ modulo $2\pi$.

Using Eq.~\eqref{eq:step-height}, the averaged arm currents are
\begin{equation}
\left\langle I_{i,\infty}\right\rangle_n
=(-1)^n\mathcal I_iJ_n(a)
\sin(\varphi_{0i}+n\psi_i).
\label{eq:two-arm-averaged-current}
\end{equation}
Maximizing their sum over the common dc phase gives
\begin{equation}
I_{n,\infty}^{\rm mw}(f)=|J_n(a)|
\sqrt{\mathcal I_1^2+\mathcal I_2^2
+2\mathcal I_1\mathcal I_2
\cos(2\pi f-n\Delta\psi)}.
\label{eq:mw-squid-envelope}
\end{equation}
The corresponding translation of the dc-flux offset is
\begin{equation}
\frac{\Delta\Phi_{\rm mw}}{\PhiQ}=\frac{n\Delta\psi}{2\pi}
\quad (\mathrm{mod}\ 1).
\label{eq:mw-flux-shift}
\end{equation}
This is a propagation-and-drive effect conditional on Eq.~\eqref{eq:dynamic-flux-balance}, not a direct shift generated by the static lapse.  Unequal locally calibrated microwave amplitudes also change the Bessel weights, but that deformation is protocol dependent.

\begin{figure}[t]
\centering
\includegraphics[width=\columnwidth]{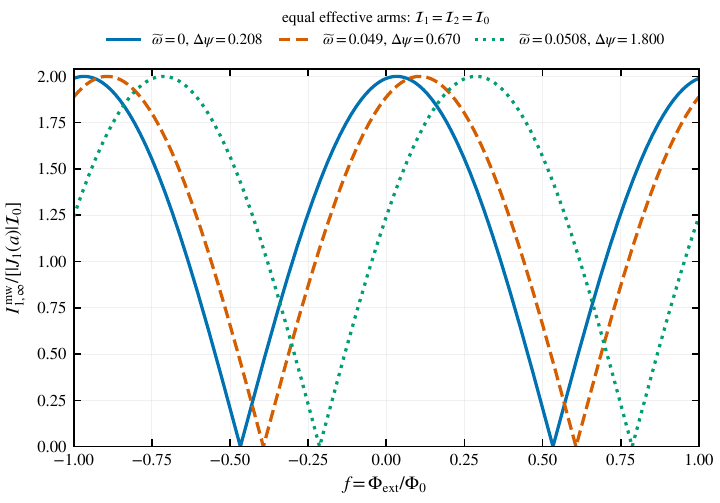}
\caption{Conditional propagation-induced translation of the normalized $n=1$ microwave-locked envelope for $\gamma=9/2$.  The curves set $\mathcal I_1=\mathcal I_2=\mathcal I_0$ and plot $I_{1,\infty}^{\rm mw}/[|J_1(a)|\mathcal I_0]$, thereby isolating phase translation from arm-amplitude imbalance.  The radii $x_++10^{-3}$ and $x_++2\times10^{-3}$ are propagation-kernel endpoints used to determine $\Delta\psi$; the paths terminate at $x=5$ and use $\Omega_\infty\rs/c=0.30$.  Holding the coordinate-radius endpoint mismatch fixed enhances the phase shift over the displayed nonextremal range, but the fixed-offset limit ultimately crosses over to the pole-controlled extremal kernel.  The envelope assumes the matching ac loop flux in Eq.~\eqref{eq:required-ac-loop-flux}.}
\label{fig:squid-mw}
\end{figure}

To isolate the phase translation from arm-amplitude imbalance, Fig.~\ref{fig:squid-mw} sets $\mathcal I_1=\mathcal I_2\equiv\mathcal I_0$ and plots $I_{1,\infty}^{\rm mw}/[|J_1(a)|\mathcal I_0]$.  For the coordinate-radius offsets used in the figure, the exact propagation phases are $\Delta\psi=0.20824$, $0.67047$, and $1.80037$ for $\wtil=0$, $0.049$, and $0.0508$, respectively.  These correspond to translations of $0.0331$, $0.1067$, and $0.2865$ flux periods.  Equal effective arm amplitudes must be imposed by circuit calibration if the propagation endpoints are also the junction radii.  The comparison holds the coordinate offsets fixed; it does not hold the proper radial separation fixed.  Over the displayed parameter range the simple-horizon logarithm remains important, whereas at fixed offsets an arbitrarily close approach to extremality eventually crosses over according to Eq.~\eqref{eq:uniform-differential-delay}.

For finite geometric self-inductance,
\begin{equation}
\Phi_{\rm tot}(t)=\Phi_{\rm ext}(t)+L_sI_{\rm circ}(t),
\label{eq:finite-inductance-flux}
\end{equation}
and Eq.~\eqref{eq:dynamic-flux-balance} must be solved together with the two RCSJ equations.  If the kinetic inductance is appreciable, the bulk gauge-invariant condensate momentum must also be retained in the fluxoid relation.  A distributed loop spanning a significant lapse gradient therefore requires a spatially resolved electrodynamic and condensate model.  In particular, neither the circulating current nor a time-dependent total flux can be replaced by the total transport current.

\section{Tolman--KMS shunt noise and a universal
horizon fluctuation--delay relation}
\label{sec:noise-delay}

The preceding analysis concerns mean Josephson transport.  A resistively
shunted junction also contains equilibrium current fluctuations fixed by
the fluctuation--dissipation theorem
\cite{CallenWelton1951,Clerk2010}.  This permits a quantum-noise extension
of the redshifted RCSJ dictionary and leads to a parameter-independent
relation between the near-horizon null delay and the equilibrium
shunt noise.

We use a two-sided symmetrized spectrum in angular frequency.  For the
local Norton current-noise source of an ideal Ohmic shunt, whose resistance
is taken to be frequency-independent over the band of interest

\begin{equation}
S_{I,{\rm p}}^{\rm sym}(\omega)
\equiv
\frac{1}{2}
\int_{-\infty}^{\infty}\dif\Delta\tau\,
e^{\imag\omega\Delta\tau}
\,
\left\langle
\left\{
\delta I_{\rm N}^{\rm p}(\Delta\tau),
\delta I_{\rm N}^{\rm p}(0)
\right\}
\right\rangle ,
\label{eq:local-current-noise-definition}
\end{equation}
the quantum fluctuation--dissipation relation gives
\begin{equation}
S_{I,{\rm p}}^{\rm sym}(\omega)
=
\frac{\hbar|\omega|}{R}
\coth\left(
\frac{\hbar|\omega|}{2k_{\rm B}T_{\rm p}}
\right).
\label{eq:local-current-noise-fdt}
\end{equation}
Its low-frequency limit is
$S_{I,{\rm p}}^{\rm sym}(0)=2k_{\rm B}T_{\rm p}/R$.

For a compact static junction,
\begin{equation}
\delta I_{{\rm N},\infty}(t)
=
\mathcal N_J\,
\delta I_{\rm N}^{\rm p}
(\tau=\mathcal N_J t).
\label{eq:noise-current-map}
\end{equation}
We denote the angular frequency conjugate to the asymptotically normalized
Killing time by $\omega_\infty$ and adopt the Fourier convention
\begin{equation}
\begin{aligned}
\delta X(\omega_\infty)
&\equiv\int_{-\infty}^{\infty}\dif t\,
\ee^{\imag\omega_\infty t}\delta X(t),\\
\partial_t&\longmapsto-\imag\omega_\infty.
\end{aligned}
\label{eq:killing-time-fourier-convention}
\end{equation}
This notation distinguishes a generic Killing-time spectral frequency from
both the gravitational parameter $\Omega$ and the microwave frequency
$\Omega_\infty$.
The Fourier rescaling therefore produces
\begin{equation}
S_{I,\infty}^{\rm sym}(\omega_\infty)
=
\mathcal N_J
S_{I,{\rm p}}^{\rm sym}
\left(\frac{\omega_\infty}{\mathcal N_J}\right).
\label{eq:noise-spectrum-map}
\end{equation}
In the nonextremal Hartle--Hawking state,
$T_{\rm p}=T_H/\N_J$, where $T_H$ is the asymptotic Hawking
temperature $T_\infty=T_H=\frac{\hbar\kappa_+}{2\pi c k_{\rm B}}$
\cite{Tolman:1930zza,Tolman:1930ona,Hawking:1975,HartleHawking:1976}.
Substitution into Eq.~\eqref{eq:noise-spectrum-map} yields
\begin{equation}
S_{I,\infty}^{\rm sym}(\omega_\infty)
=
\frac{\hbar|\omega_\infty|}{R}
\coth\left(
\frac{\hbar|\omega_\infty|}{2k_{\rm B}T_H}
\right).
\label{eq:HH-noise-infinity}
\end{equation}
All lapse factors have cancelled.  Here and below,
$S_{I,\infty}^{\rm sym}$ denotes the asymptotic-time-referred Norton
source spectrum, not the total current-noise spectrum after filtering by
an external circuit.  Thus shunts with equal local resistance have the
same referred source spectrum at different static radii.  More generally,
\begin{equation}
R_i S_{I_i,\infty}^{\rm sym}(\omega_\infty)
=
R_j S_{I_j,\infty}^{\rm sym}(\omega_\infty)
\label{eq:cross-radius-noise-null-test}
\end{equation}
for any two locally equilibrated Ohmic shunts in the same
Hartle--Hawking state.

The stochastic asymptotic RCSJ equation is
\begin{equation}
\frac{\hbar C}{q\mathcal N_J}\ddot\varphi
+\frac{\hbar}{qR}\dot\varphi
+\mathcal N_J I_c^{\rm p}\sin\varphi
=
I_\infty+\delta I_{{\rm N},\infty}(t).
\label{eq:stochastic-asymptotic-rcsj}
\end{equation}
Linearizing about a stable stationary solution $\varphi_0$, satisfying
$I_\infty=\N_J I_c^{\rm p}\sin\varphi_0$ and
$\cos\varphi_0>0$, we write
$\varphi(t)=\varphi_0+\delta\varphi(t)$ and define the small-signal
phase--current susceptibility by
\begin{equation}
\delta\varphi(\omega_\infty)
=\chi_{\varphi I}(\omega_\infty)
\delta I_{{\rm N},\infty}(\omega_\infty).
\label{eq:phase-current-response-definition}
\end{equation}
The linearized stochastic RCSJ equation then gives
\begin{equation}
\chi_{\varphi I}(\omega_\infty)
=
\left[
\mathcal N_J I_c^{\rm p}\cos\varphi_0
-\frac{\hbar C}{q\mathcal N_J}\omega_\infty^2
-\imag\frac{\hbar\omega_\infty}{qR}
\right]^{-1},
\label{eq:phase-current-susceptibility}
\end{equation}
which has units of inverse current and satisfies
$\chi_{\varphi I}(-\omega_\infty)
=\chi_{\varphi I}^{*}(\omega_\infty)$.  Consequently,
\begin{equation}
S_{\varphi,\infty}^{\rm sym}(\omega_\infty)
=
|\chi_{\varphi I}(\omega_\infty)|^2
S_{I,\infty}^{\rm sym}(\omega_\infty).
\label{eq:phase-noise-spectrum}
\end{equation}
Equation~\eqref{eq:phase-noise-spectrum} predicts a damped,
bias-dependent Josephson-plasma response.  Its undamped small-signal
frequency is
$\omega_{0,\infty}=\omega_{p,\infty}\sqrt{\cos\varphi_0}$; the location
of the maximum of the full phase-noise spectrum also depends on damping and
on the frequency dependence of the Norton source.  The Killing-frequency
noise temperature remains independent of the static junction radius.

A second consequence follows by combining the noise spectrum with the
near-horizon propagation kernel.  Define the coefficient of the local
logarithmic null delay by
\begin{equation}
\mathcal A_H
\equiv
\lim_{\delta\rightarrow0^+}
\frac{\partial\Delta t}
{\partial[-\ln\delta]},
\qquad
\delta=x_{\rm em}-x_+.
\label{eq:horizon-delay-coefficient-definition}
\end{equation}
Equation~\eqref{eq:near-horizon-time} gives
\begin{equation}
\mathcal A_H
=
\frac{\rs}{2c\bar\kappa_+}
=
\frac{c}{2\kappa_+}.
\label{eq:horizon-delay-coefficient}
\end{equation}
This coefficient is invariant under regular reparametrizations of the
horizon offset: if $\delta'=C\delta+\Order(\delta^2)$ with $C>0$, then
the derivative in Eq.~\eqref{eq:horizon-delay-coefficient-definition}
has the same limiting value.  Its normalization is fixed by the Killing
time normalized at infinity.  On the other hand, the zero-frequency limit of
Eq.~\eqref{eq:HH-noise-infinity} is
\begin{equation}
R S_{I,\infty}^{\rm sym}(0)
=
2k_{\rm B}T_H
=
\frac{\hbar\kappa_+}{\pi c}.
\label{eq:zero-frequency-HH-noise}
\end{equation}
Eliminating the surface gravity between
Eqs.~\eqref{eq:horizon-delay-coefficient} and
\eqref{eq:zero-frequency-HH-noise} yields
\begin{equation}
\mathcal A_H R S_{I,\infty}^{\rm sym}(0)
=
\frac{\hbar}{2\pi}.
\label{eq:universal-fluctuation-delay-relation}
\end{equation}
Equivalently, for the Thevenin voltage-noise spectrum
$S_{V,\infty}^{\rm sym}=R^2S_{I,\infty}^{\rm sym}$,
\begin{equation}
\mathcal A_H
\frac{S_{V,\infty}^{\rm sym}(0)}{R}
=
\frac{\hbar}{2\pi}.
\label{eq:voltage-fluctuation-delay-relation}
\end{equation}
The numerical prefactor in these equations corresponds to the explicit
two-sided angular-frequency convention in
Eq.~\eqref{eq:local-current-noise-definition}.

The finite-frequency generalization is
\begin{equation}
\mathcal A_H R
S_{I,\infty}^{\rm sym}(\omega_\infty)
=
\frac{\hbar}{2\pi}\,
y\coth y,
\qquad
y\equiv\frac{\pi c|\omega_\infty|}{\kappa_+}.
\label{eq:finite-frequency-fluctuation-delay}
\end{equation}
The universal constant in
Eq.~\eqref{eq:universal-fluctuation-delay-relation}  is recovered for
$y\ll1$.  At fixed nonzero frequency and $\kappa_+\rightarrow0$,
zero-point noise dominates and the low-frequency and extremal limits
do not commute.

Near extremality,
$\kappa_+=(c^2/\rs)C_\kappa\sqrt{\delta\widetilde\omega}
+\mathcal O(\delta\widetilde\omega)$, and therefore
\begin{align}
R S_{I,\infty}^{\rm sym}(0)
&=
\frac{\hbar c}{\pi\rs}
C_\kappa\sqrt{\delta\widetilde\omega}
+\mathcal O(\delta\widetilde\omega),
\label{eq:near-ext-noise}\\
\mathcal A_H
&=
\frac{\rs}{2cC_\kappa}
\frac{1}{\sqrt{\delta\widetilde\omega}}
+\mathcal O(1).
\label{eq:near-ext-delay-coefficient}
\end{align}
Thus the suppression of the low-frequency Hartle--Hawking noise exactly
compensates the enhancement of the local logarithmic delay coefficient.

The same lapse cancellation has a KMS interpretation.  The symmetrized
spectrum above characterizes equilibrium fluctuations, whereas excitation
and relaxation rates are governed by the unsymmetrized bath spectrum.
Quantizing the small-amplitude Josephson-plasma mode gives
$\omega_{p,\infty}=\N_J\omega_{p,{\rm p}}$.  In the weak-coupling,
linear-response regime, local KMS detailed balance gives
\begin{equation}
\frac{\Gamma_\uparrow}{\Gamma_\downarrow}
=
\exp\left(
-\frac{\hbar\omega_{p,{\rm p}}}{k_{\rm B}T_{\rm p}}
\right)
=
\exp\left(
-\frac{\hbar\omega_{p,\infty}}{k_{\rm B}T_H}
\right).
\label{eq:plasma-mode-kms}
\end{equation}
When the transition frequency is referred to Killing time, the rate ratio
therefore returns the asymptotic Hawking temperature independently of the
static junction radius.

Equations~\eqref{eq:HH-noise-infinity}--\eqref{eq:plasma-mode-kms}
assume a nonextremal Hartle--Hawking state, an ideal Ohmic shunt, local
equilibration, and operation outside the superconducting exclusion layer,
$\rho>\rho_c$, with $\rho_c$ defined in
Eq.~\eqref{eq:thermal-exclusion-distance}.  The parameter-free
zero-frequency relation refers to the unfiltered Norton source spectrum;
a measured terminal spectrum must include the susceptibility of the full
circuit.  These results do not apply unchanged in the Unruh state, where
there is no global KMS condition and the appropriate observable is a
state-dependent detector response.  Moreover, $\mathcal A_H$ is the
coefficient of the local Rindler logarithm; the total delay at fixed
coordinate endpoints must be evaluated with the uniform
Rindler-to-$\mathrm{AdS}_2$ crossover when extremality is approached.

\section{Validity, thermal exclusion, and observability}\label{sec:validity}

A static junction of proper thickness $L$ must satisfy
\begin{equation}
L\ll\xi_{L,R},
\qquad
L\ll\ell_\N,
\qquad
L\ll R_{\rm surf},
\qquad
L\ll R_c,
\label{eq:validity-lengths}
\end{equation}
where $R_{\rm surf}$ is the proper curvature radius of the barrier interfaces and $R_c$ is a local ambient-curvature scale.  The condensate and electromagnetic stress tensors must remain negligible relative to the source of the background.  Near a simple horizon, Eq. \eqref{eq:Rindler-lapse} implies
\begin{equation}
\ell_\N\simeq\rho,
\label{eq:lapse-length-near-horizon}
\end{equation}
so the short-junction approximation itself requires $L\ll\rho$.

The proper acceleration of a static terminal is
\begin{equation}
a_{\rm stat}=\frac{c^2}{2\rs}\frac{|F_x|}{\sqrt F},
\label{eq:static-acceleration}
\end{equation}
which diverges as $F^{-1/2}$.  Free fall removes the support acceleration but replaces the static clock map by the received factor in Eq. \eqref{eq:received-redshift} and introduces tidal and motional-emf constraints.

\subsection{Universal thermal exclusion layer}

In static thermal equilibrium,
\begin{equation}
T_{\rm p}(r)=\frac{T_\infty}{\N(r)}.
\label{eq:Tolman-temperature}
\end{equation}
For a nonextremal black hole in the Hartle--Hawking state \cite{Hawking:1975,HartleHawking:1976},
\begin{equation}
T_\infty=T_H=\frac{\hbar\kappa_+}{2\pi c k_{\rm B}}.
\label{eq:hawking-temperature}
\end{equation}
Combining this with the Rindler lapse \eqref{eq:Rindler-lapse} eliminates both the surface gravity and the black-hole mass:
\begin{equation}
T_{\rm p}(\rho)\simeq\frac{\hbar c}{2\pi k_{\rm B}\rho}.
\label{eq:universal-local-temperature}
\end{equation}
A superconductor with local critical temperature $T_c$ can remain in equilibrium only for
\begin{equation}
\rho>\rho_c,
\qquad
\rho_c=\frac{\hbar c}{2\pi k_{\rm B}T_c}.
\label{eq:thermal-exclusion-distance}
\end{equation}
Numerically,
\begin{equation}
\rho_c=0.364~{\rm mm}\left(\frac{1~{\rm K}}{T_c}\right).
\label{eq:thermal-exclusion-numeric}
\end{equation}
The leading exclusion distance is universal for every nonextremal horizon with a regular Rindler region.  Together with Eq. \eqref{eq:lapse-length-near-horizon}, a compact equilibrium device must satisfy
\begin{equation}
\rho>\rho_c,
\qquad
\frac{L}{\rho}\ll1.
\label{eq:combined-near-horizon-bound}
\end{equation}
The Hartle--Hawking assumption is essential: an evaporating black hole in the Unruh state \cite{Unruh:1976} is not in global static equilibrium, and a detector-response calculation must replace the Tolman argument.  Equations~\eqref{eq:universal-local-temperature}--\eqref{eq:combined-near-horizon-bound} also require $\rho_c$ to lie within the regular nonextremal Rindler domain.  At exact extremality, substituting $T_H=0$ into the nonextremal result is only formal: the extremal state and throat are not obtained by a uniform near-horizon limit, and the limits $\kappa_+\to0$ and $\rho\to0$ need not commute.  The extremal thermal problem must therefore be treated separately.

\begin{figure}[t]
\centering
\includegraphics[width=\columnwidth]{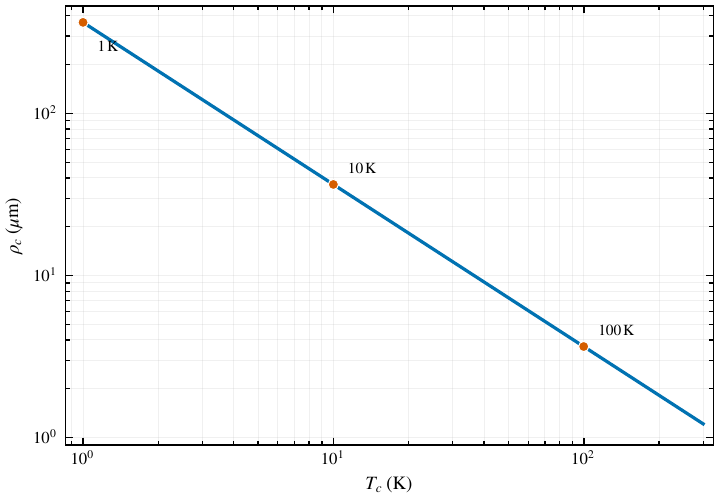}
\caption{Leading proper-distance superconducting exclusion scale in a Hartle--Hawking bath near a regular nonextremal Rindler horizon.  At this order, $\rho_c=\hbar c/(2\pi k_{\rm B}T_c)$ depends only on the local critical temperature and not on $M$, $\kappa_+$, or $\wtil$.}
\label{fig:thermal-exclusion}
\end{figure}

Figure~\ref{fig:thermal-exclusion} displays the inverse-temperature scaling of Eq.~\eqref{eq:thermal-exclusion-distance}.  The exclusion distances are $364.4~\mu{\rm m}$, $36.44~\mu{\rm m}$, and $3.644~\mu{\rm m}$ for $T_c=1$, $10$, and $100~{\rm K}$, respectively.  These values are independent of the black-hole parameters only at leading Rindler order in Hartle--Hawking equilibrium.

To display the physical scale explicitly, write
$\Omega=\eta_\Omega\ell_{\rm P}^2$, where
$\ell_{\rm P}=\sqrt{\hbar G_0/c^3}$ is the Planck length,
$m_{\rm P}=\sqrt{\hbar c/G_0}$ is the Planck mass, and $\eta_\Omega$ is
dimensionless.  Then
\begin{equation}
\begin{aligned}
|\wtil|
&=|\eta_\Omega|\left(\frac{\ell_{\rm P}}{\rs}\right)^2
=\frac{|\eta_\Omega|}{4}
\left(\frac{m_{\rm P}}{M}\right)^2\\
&=2.99\times10^{-77}|\eta_\Omega|
\left(\frac{M_\odot}{M}\right)^2.
\end{aligned}
\label{eq:running-observability-scale}
\end{equation}
Consequently, for $|\eta_\Omega|=\Order(1)$, the scale-dependent lapse,
shadow, and propagation corrections are negligible for stellar-mass and
supermassive black holes.  Primordial origin does not change this mass
scaling.  For example, a primordial black hole with
$M=5\times10^{11}\,{\rm kg}$, of the order of the mass whose standard
semiclassical Hawking lifetime is comparable to the age of the Universe
\cite{Hawking:1975}, has
\begin{equation}
|\wtil|\simeq4.7\times10^{-40}|\eta_\Omega|.
\label{eq:pbh-running-scale}
\end{equation}
Thus long-lived primordial black holes remain deep in the weak-running
regime.  Lighter evaporating objects acquire larger corrections as $M^{-2}$,
but $|\wtil|$ becomes appreciable only during a terminal Planck-mass phase or
for a hypothetical Planck-scale remnant.  In that regime the stationary
semiclassical background, equilibrium-state description, and material
test-junction approximation used here are not simultaneously controlled.
Moreover, the above reference mass has a Schwarzschild Hawking temperature
$T_H\simeq2.5\times10^{11}\,{\rm K}$, precluding an ordinary superconducting device in
Hartle--Hawking equilibrium.  Primordial black holes therefore do not provide
a realizable material enhancement of the effects studied here, although their
terminal quantum-gravity regime is a distinct problem.  More specifically,
the positive-running near-extremal examples with $\gamma=9/2$ require
\begin{equation}
\sqrt{\Omega}=\sqrt{\wtil_e}\,\rs=0.226\,\rs.
\label{eq:near-ext-running-scale}
\end{equation}
If $\Omega=\ell_{\rm P}^2$, this corresponds to $\rs=4.43\,\ell_{\rm P}$ and $M=2.21\,m_{\rm P}$.  The near-extremal curves should therefore be read as dimensionless benchmarks for an enhanced-running or analogue realization, not as order-unity predictions of Planck-scale running around an astrophysical black hole.  The Hartle--Hawking proper-distance bound is not suppressed by $\wtil$, but it relies on the equilibrium-state assumption stated above.  The most robust formal targets are dimensionless ratios, differential phases, and correlated observables: the one-lapse static transfer, the distinction between coordinate and received infall factors, the fixed dc lobe centers at negligible total inductance, the local near-extremal logarithmic propagation coefficient together with its fixed-offset extremal crossover, the Tolman--KMS cross-radius shunt-noise null test and fluctuation--delay relation, the universal thermal-distance bound, and the leading shadow--Josephson consistency relation.

\section{Shadow--Josephson consistency relation}
\label{sec:shadow-consistency}

The same running parameter that controls the Josephson clock transfer also
changes the unstable circular null orbit.  For the metric in
Eq.~\eqref{eq:metric-physical}, a photon sphere satisfies \cite{Synge:1966,Perlick:2021}
\begin{equation}
 x F_x(x)-2F(x)=0.
 \label{eq:photon-sphere-condition}
\end{equation}
Using Eq.~\eqref{eq:F-D}, this condition becomes
\begin{multline}
2x^6-3x^5+4\wtil x^4+(2\gamma-1)\wtil x^3
+2\wtil^2x^2 \\
+2\gamma\wtil^2x+\frac{\gamma^2}{2}\wtil^2=0.
\label{eq:photon-sphere-polynomial}
\end{multline}
A circular null orbit contributes to the shadow edge only if it lies outside the outer horizon and is unstable.  The instability condition is
\begin{equation}
\begin{aligned}
\left.
\frac{\dif^2}{\dif x^2}\left(\frac{F(x)}{x^2}\right)
\right|_{x=x_{\rm ph}}&<0,\\
&\Longleftrightarrow\quad
2F(x_{\rm ph})-x_{\rm ph}^2F_{xx}(x_{\rm ph})>0.
\end{aligned}
\label{eq:photon-sphere-instability}
\end{equation}
The physical root must satisfy both $x_{\rm ph}>x_+$ and Eq.~\eqref{eq:photon-sphere-instability}.  The perturbative outer root continuously connected to the Schwarzschild value $x_{\rm ph}=3/2$ satisfies these conditions.  For small running,
\begin{equation}
 x_{\rm ph}=\frac32-\frac{10+4\gamma}{9}\wtil
 +\Order(\wtil^2).
 \label{eq:photon-sphere-small-running}
\end{equation}
The critical impact parameter measured at infinity is
\begin{equation}
 b_{\rm sh}=\frac{r_s x_{\rm ph}}{\sqrt{F(x_{\rm ph})}},
 \label{eq:shadow-impact-parameter}
\end{equation}
which gives
\begin{equation}
 \frac{b_{\rm sh}}{r_s}=\frac{3\sqrt3}{2}
 \left[1-\frac{4(\gamma+3)}{27}\wtil
 +\Order(\wtil^2)\right].
 \label{eq:shadow-small-running}
\end{equation}

Consider a static junction satisfying $x_J>x_+(\wtil)$.  For the fixed-$x_J$ expansion about Schwarzschild we additionally take $x_J>1$ and require $\epsilon_{\wtil}(x_J)\ll1$ as defined in Eq.~\eqref{eq:weak-running-validity}.  Under these conditions, any proper observable that acquires one lapse factor---for example $I_{c,\infty}$ or $\omega_{p,\infty}$ at fixed proper junction parameters---has the fractional shift
\begin{equation}
 \frac{\delta\mathcal O_J}{\mathcal O_{J,\rm Schw}}
 =\frac{\wtil}{2F_{\rm Schw}(x_J)}
 \left(\frac1{x_J^3}+\frac{\gamma}{2x_J^4}\right)
 +\Order(\wtil^2).
 \label{eq:one-lapse-small-running-general}
\end{equation}
Eliminating $\wtil$ between Eqs.~\eqref{eq:shadow-small-running} and
\eqref{eq:one-lapse-small-running-general} yields
\begin{multline}
 \frac{\delta\mathcal O_J}{\mathcal O_{J,\rm Schw}}
 =-\frac{27}{8(\gamma+3)F_{\rm Schw}(x_J)}
 \left(\frac1{x_J^3}+\frac{\gamma}{2x_J^4}\right) \\
 \times\frac{\delta b_{\rm sh}}{b_{\rm sh,Schw}}
 +\Order\!\left[\left(\frac{\delta b_{\rm sh}}
 {b_{\rm sh,Schw}}\right)^2\right].
 \label{eq:shadow-josephson-consistency}
\end{multline}
Equation~\eqref{eq:shadow-josephson-consistency} is a leading-order cross-channel consistency relation: once $\gamma$ and the fixed junction radius are specified, the running parameter no longer appears.  Its use near the shifted horizon requires the exact metric rather than the nonuniform small-$\wtil$ expansion.  The relation is primarily an internal model-consistency test or an analogue-system target; it is not, by itself, a proposal to place a material junction near an astrophysical horizon.  In an astronomical application, the inferred shadow shift must also be marginalized over the mass-to-distance ratio and image-model systematics, while the Josephson side requires an independently specified device and clock-comparison protocol.

A null interferometric result can be converted into a bound on the running
parameter only after the measured feature and its protocol are specified.
In the zero-total-inductance dc envelope, a static lapse imbalance changes the lobe
amplitudes but not their centers; therefore, a dc lobe-center uncertainty alone
does not constrain $\wtil$.  The conditional microwave translation in
Eq.~\eqref{eq:mw-flux-shift} does provide a phase-sensitive constraint.  Let
$\sigma_f\equiv\sigma_\Phi/\PhiQ$ be the one-standard-deviation uncertainty of
the dimensionless lobe-center position and $N_\sigma$ the desired confidence
multiplier.  Linearizing about a reference $\wtil_0$ gives
\begin{equation}
|\wtil-\wtil_0|
\lesssim
\frac{2\pi N_\sigma\sigma_f}
{|n|\Omega_\infty
\left|\partial_{\wtil}(\Delta t_1-\Delta t_2)
\right|_{\wtil_0}}.
\label{eq:exact-interferometric-bound}
\end{equation}
For fixed exterior endpoints, the derivative of each exact propagation time
can be evaluated without finite differencing:
\begin{equation}
\partial_{\wtil}\Delta t_{ab}
=-\frac{\rs}{c}\int_{x_a}^{x_b}
\frac{x^2(x+\gamma/2)}{\Pp(x;\wtil)^2}\,\dif x.
\label{eq:propagation-derivative-bound}
\end{equation}
In the weak-running regime about Schwarzschild, define
$\widehat\Omega_\infty\equiv\Omega_\infty\rs/c$ and
\begin{equation}
\Delta\mathcal K_{12}
\equiv[\mathcal K(x_{b1})-\mathcal K(x_{a1})]
-[\mathcal K(x_{b2})-\mathcal K(x_{a2})].
\label{eq:delta-K-paths}
\end{equation}
Equation~\eqref{eq:exact-interferometric-bound} then reduces to
\begin{equation}
|\wtil|\lesssim
\frac{2\pi N_\sigma\sigma_f}
{|n|\widehat\Omega_\infty|\Delta\mathcal K_{12}|}.
\label{eq:weak-interferometric-bound}
\end{equation}
Independently, a one-lapse Josephson observable measured at a fixed
$x_J>1$ with fractional uncertainty $\sigma_J$ gives
\begin{equation}
|\wtil|\lesssim
\frac{2F_{\rm Schw}(x_J)N_\sigma\sigma_J}
{x_J^{-3}+\gamma/(2x_J^4)}.
\label{eq:one-lapse-running-bound}
\end{equation}
The corresponding dimensional bound is $|\Omega|\leq\rs^2|\wtil|$.
These sensitivity relations assume calibrated junction parameters, path
geometry, mass, and radii.  If an endpoint follows
$x_+(\wtil)+\delta$, its endpoint derivative must be included; near the shifted
horizon one must differentiate the exact kernel rather than use
Eq.~\eqref{eq:weak-interferometric-bound}.  A numerical bound is therefore not
meaningful until $\sigma_f$ (or $\sigma_J$), the apparatus geometry, and the
phase-systematics budget are specified.

\section{Conclusion}\label{sec:conclusion}

We developed a covariant Josephson framework for the scale-dependent exterior generated by an Oppenheimer--Snyder-like collapse.  The rational blackening function admits an analytic classification of horizons and singular shells.  For $\gamma>0$, its positive-running center is curvature-finite, with finite orthonormal-frame Riemann components and polynomial invariants, but the nonzero linear radial corrections show that this does not establish an analytic Cartesian extension or geodesic completeness.  The extremal branch develops an $\mathrm{AdS}_2\times S^2$ throat and a surface gravity that vanishes as $\sqrt{\wtil_e-\wtil}$.

Within the minimally coupled test-condensate approximation, local weak-link physics retains the standard Ginzburg--Landau current-phase relation when all material parameters are defined and held fixed in the orthonormal frame.  Static quantities referred to Killing time acquire a controlled lapse dependence: the ac frequency and charge current carry one factor of $\sqrt F$, and power carries $F$.  The RCSJ equation shows that the plasma frequency redshifts by the lapse while the Stewart--McCumber parameter remains invariant for a compact static junction.

For a collapsing terminal, $F/\mathcal E$ is the conversion between proper time and the exterior coordinate time at emission.  The frequency of an outgoing signal received at infinity instead carries $\mathcal E-\sqrt{\mathcal E^2-F}$.  This distinction preserves the leading $F$ suppression but changes the normalization and the natural late-time law: the received signal decays as $\exp(-\kappa_+u/c)$ for a simple horizon and as $u^{-2}$ at extremality.

The exact radial-null travel-time kernel has a local logarithmic coefficient proportional to $1/\kappa_+$ near a simple horizon.  This enhancement applies while the relevant endpoints remain inside the shrinking Rindler cap; at fixed nonzero offsets, the approach to extremality crosses over to the pole-controlled extremal kernel, so the $\kappa_+\to0$ and endpoint limits do not commute.  Its use as a microwave phase law further assumes the eikonal and narrow-band conditions stated in Eq.~\eqref{eq:geometric-optics-validity}.  Static lapse imbalance deforms a dc SQUID envelope without translating its lobe centers in the zero-total-inductance limit.  Under microwave drive, a differential propagation phase translates the envelope only when the associated time-dependent loop flux satisfies the dynamical fluxoid constraint; it is not an independent static-lapse shift.  A null microwave lobe-shift measurement or a one-lapse amplitude measurement yields the explicit error-propagation bounds in Eqs.~\eqref{eq:exact-interferometric-bound} and \eqref{eq:one-lapse-running-bound}, once the experimental geometry and systematics are specified.  A Hartle--Hawking equilibrium bath excludes superconductivity from the leading nonextremal layer $\rho<\hbar c/(2\pi k_{\rm B}T_c)$, provided that this layer lies within the Rindler domain.  Finally, the leading shadow and one-lapse shifts obey the parameter-eliminated consistency relation \eqref{eq:shadow-josephson-consistency}.

For the asymptotic-time-referred Norton source spectrum in a nonextremal
Hartle--Hawking state, the inverse-surface-gravity enhancement of the local
null-delay coefficient is exactly compensated by the suppression of the
low-frequency shunt noise.  Their product is $\hbar/(2\pi)$ in the
specified two-sided angular-frequency convention, providing a
parameter-free fluctuation--delay consistency test.

Natural next steps are a finite-inductance two-phase RCSJ treatment with a self-consistent microwave network, rotating scale-dependent backgrounds, and a detector-level calculation in the Unruh state for an evaporating black hole.

\begin{acknowledgments}
R.P., G.L., and A.\"O. acknowledge networking support from COST Actions CA21106, CA22113, CA21136, CA23130, and CA23115, funded by COST (European Cooperation in Science and Technology).  A.\"O. also acknowledges support from EMU, T\"UB\.{I}TAK, ULAKB\.{I}M (T\"urkiye), and SCOAP$^3$ (Switzerland).
\end{acknowledgments}

\appendix

\section{Algebraic identities}\label{app:algebra}

Differentiating Eq. \eqref{eq:F-D} gives
\begin{equation}
F_x=\frac{x^4-\wtil x^2-\gamma\wtil x}{\D^2}.
\label{eq:Fx-general}
\end{equation}
At a horizon, $\D(x_h)=x_h^2$ and $\Pp(x_h)=0$, so
\begin{equation}
F_x(x_h)=3-\frac2{x_h}+\frac{\wtil}{x_h^2}
=1-\frac{\wtil}{x_h^2}-\frac{\gamma\wtil}{x_h^3}.
\label{eq:Fx-horizon-identities}
\end{equation}
The extremal identities are
\begin{equation}
\wtil_e=2x_e-3x_e^2,
\qquad
2x_e^2-\left(1-\frac{3\gamma}{2}\right)x_e-\gamma=0,
\label{eq:extremal-identities}
\end{equation}
and
\begin{equation}
F_{xx}(x_e)=\frac{\Pp''(x_e)}{x_e^2}=\frac{6x_e-2}{x_e^2}.
\label{eq:Fxx-proof}
\end{equation}
The weak-running expansion follows from
\begin{equation}
\frac{x^2}{\D(x)}=\frac1x
\left[1+\wtil\left(\frac1{x^2}+\frac{\gamma}{2x^3}\right)\right]^{-1}.
\label{eq:weak-expansion-start}
\end{equation}

\section{Propagation kernels}\label{app:propagation}

For simple roots, the residue identity is
\begin{equation}
A_j=\lim_{x\to x_j}\frac{x^2(x-x_j)}{\Pp(x)}
=\frac{x_j^2}{\Pp'(x_j)}.
\label{eq:residue-proof}
\end{equation}
At extremality,
\begin{equation}
\Pp(x)=(x-x_e)^2(x-x_0),
\qquad
x_0=1-2x_e.
\label{eq:extremal-factorization}
\end{equation}
The rational part decomposes as
\begin{equation}
\frac{x^2}{\Pp(x)}=
\frac{\ell_2^2}{(x-x_e)^2}
+\frac{B_e}{x-x_e}+\frac{B_0}{x-x_0},
\label{eq:extremal-partial-fraction}
\end{equation}
with
\begin{equation}
B_e=\frac{x_e(5x_e-2)}{(3x_e-1)^2},
\qquad
B_0=\frac{(2x_e-1)^2}{(3x_e-1)^2}.
\label{eq:extremal-log-coefficients}
\end{equation}
Hence the extremal tortoise kernel is
\begin{multline}
\int\frac{\dif x}{F(x)}=
 x-\frac{\ell_2^2}{x-x_e}
+B_e\ln|x-x_e|\\
+B_0\ln|x-x_0|+\mathrm{const}.
\label{eq:extremal-tortoise}
\end{multline}
The double-pole term explains the stronger extremal delay.

For the weak-running kernel,
\begin{equation}
\frac1F=\frac1{F_{\rm Schw}}
-\wtil\frac{x^{-3}+\gamma/(2x^4)}{F_{\rm Schw}^2}
+\Order(\wtil^2),
\label{eq:inverse-F-weak}
\end{equation}
where $\mathcal K'(x)=[x^{-3}+\gamma/(2x^4)]/F_{\rm Schw}^2$, as required by Eq.~\eqref{eq:inverse-F-weak}.

\section{Worldline and received redshift}\label{app:worldline}

For radial geodesic motion,
\begin{equation}
\mathcal E=F\frac{\dif t}{\dif\tau},
\qquad
\frac1{c^2}\left(\frac{\dif r}{\dif\tau}\right)^2=\mathcal E^2-F.
\label{eq:general-radial-geodesic}
\end{equation}
This immediately gives $\dif\tau/\dif t=F/\mathcal E$.  For an ingoing geodesic, $\dif r/\dif\tau=-c\sqrt{\mathcal E^2-F}$ and $\dif r_*/\dif r=1/F$, so
\begin{equation}
\frac{\dif u}{\dif\tau}
=\frac{\mathcal E}{F}-\frac1c\frac{\dif r_*}{\dif\tau}
=\frac{\mathcal E+\sqrt{\mathcal E^2-F}}{F}.
\label{eq:worldline-retarded-proof}
\end{equation}
Rationalizing its inverse gives Eq. \eqref{eq:received-redshift}.

For a time-dependent connector, differentiating Eq. \eqref{eq:gauge-phase} also produces
\begin{equation}
-\frac{q}{\hbar}\frac{\dif}{\dif t}
\int_{\mathcal C(t)}A_\mu\dif x^\mu,
\label{eq:moving-connector-term}
\end{equation}
which contains the covariant electromotive contribution omitted in the electrostatic single-junction setup.

\section{Microwave harmonic average}\label{app:shapiro}

Using
\begin{equation}
\ee^{\imag a\sin(\Omega_\infty t-\psi)}
=\sum_{m=-\infty}^{\infty}J_m(a)
\ee^{\imag m(\Omega_\infty t-\psi)},
\label{eq:jacobi-anger}
\end{equation}
the zero-frequency term at $\omega_0=n\Omega_\infty$ has $m=-n$.  Since $J_{-n}=(-1)^nJ_n$,
\begin{multline}
\left\langle\sin[\varphi_0+n\Omega_\infty t
+a\sin(\Omega_\infty t-\psi)]\right\rangle\\
=(-1)^nJ_n(a)\sin(\varphi_0+n\psi).
\label{eq:shapiro-average-proof}
\end{multline}
This proves Eq. \eqref{eq:step-height}.

\bibliography{ref}

\end{document}